\documentclass{emulateapj}
\usepackage{graphicx}
\usepackage{array}

\slugcomment{Accepted for publication in ApJ}

\shorttitle{Radio Variability in Stripe 82}
\shortauthors{Hodge et al.}

\begin{document}

\title{Millijansky Radio Variability in SDSS Stripe 82}

\author{J. A. Hodge\altaffilmark{1,2}, R. H. Becker\altaffilmark{1,3}}
\altaffiltext{1}{University of California, 1 Shields Ave, Davis, CA 95616}
\email{hodge@mpia.de}

\altaffiltext{2}{Max Planck Institute for Astronomy, K\"{o}nigstuhl 17, 69117, Heidelberg, Germany}
\altaffiltext{3}{Lawrence Livermore National Laboratory, L--413, Livermore, CA 94550}

\author{R. L. White\altaffilmark{4}}
\altaffiltext{4}{Space Telescope Science Institute, 3700 San Martin Drive, Baltimore, MD 21218}

\author{G.T. Richards\altaffilmark{5}}
\altaffiltext{5}{Drexel University, 3141 Chestnut St., Philadelphia, PA 19104}


\begin{abstract} 
We report on a blind survey for extragalactic radio variability that was carried out by comparing two epochs of data from the FIRST survey with a third epoch from a new 1.4 GHz survey of SDSS Stripe 82.  The three epochs are spaced seven years apart and have an overlapping area of 60 deg$^{2}$.  We uncover 89 variable sources down to the millijansky level, 75 of which are newly--identified, and we find no evidence for transient phenomena. This new sample of variable sources allows us to infer an upper limit to the mean characteristic timescale of AGN radio variability of 14 years.  We find that only 1\% of extragalactic sources have fractional variability $f_{var}>3$, while 44\% of Galactic sources vary by this much. The variable sample contains a larger fraction of quasars than a comparable non--variable control sample, though the majority of the variable sources appear to be extended galaxies in the optical. This implies that either quasars are not the dominant contributor to the variability of the sample, or that the deep optical data allow us to detect the host galaxies of some low--z quasars. We use the new, higher resolution data to report on the morphology of the variable sources. Finally, we show that the fraction of sources that are variable remains constant or increases at low flux densities. This may imply that next generation radio surveys with telescopes like the Australian Square Kilometer Array Pathfinder and MeerKAT will see a constant or even increasing fraction of variable sources down into the submillijansky regime.

\end{abstract}


\section{Introduction}
\label{intro}

Radio sources with variable continuum emission cover a broad range of interesting astrophysical phenomena, from flaring stars to accreting supermassive black holes.    
Within our Galaxy, we see radio variability from pulsars, magnetars, microquasars, brown dwarfs, cataclysmic variables, and many low mass stars of spectral types K and M \citep{Berger:2006p629}.  Outside the Milky Way, we see exotic phenomena like Gamma--Ray Bursts, radio supernovae \citep{Weiler:2002p387}, and blazars.  
Moreover, as the time domain is still relatively unexplored, new classes of sources continue to be discovered, such as the mysterious rotating radio transients \citep[RRAT;][]{McLaughlin:2006p817}.

Several upcoming surveys are citing the untapped potential of the time domain and making variability a priority.  
The Australian Square Kilometer Array Pathfinder \citep[ASKAP;][]{Johnston:2008p151} is one upcoming telescope with variability as a priority.  Under development in a remote region of Western Australia, it will consist of 36 antennas with wide field of view phased array feeds.  The ASKAP Survey for Variables and Slow Transients \citep[VAST;][]{2013PASA...30....6M} is being planned with the goal of characterizing the radio transient and variable sky.  
Apertif is a project to upgrade the Westerbork Synthesis Radio Telescope (WSRT), increasing its field of view by a factor of 25 and facilitating the detection of variables \citep{2010iska.meetE..43O}.
Transient sources are one of the prime science drivers for MeerKAT, which is being built in the Northern Cape of South Africa and will be the largest and most sensitive radio telescope in the Southern Hemisphere upon its completion \citep{2012AfrSk..16..101B}.
The Karl G. Jansky Very Large Array (VLA) represents an upgrade to the original VLA that is more than 10 times more sensitive and will allow the quick detection of varible and transient sources. 
The Murchison Widefield Array \citep[MWA;][]{2009IEEEP..97.1497L} will be sensitive to transient radio events in the range 80--300 MHz, and LOFAR \citep{2009ASPC..407..318H} has made the search for low--frequency (10--250 MHz) transients one if its key science goals. Further in the future, the Square Kilometer Array \citep[SKA;][]{2000pras.conf.....V} is expected to contribute significantly to the transient parameter space.  

\begin{figure*}   
\centering
\includegraphics[scale=0.75]{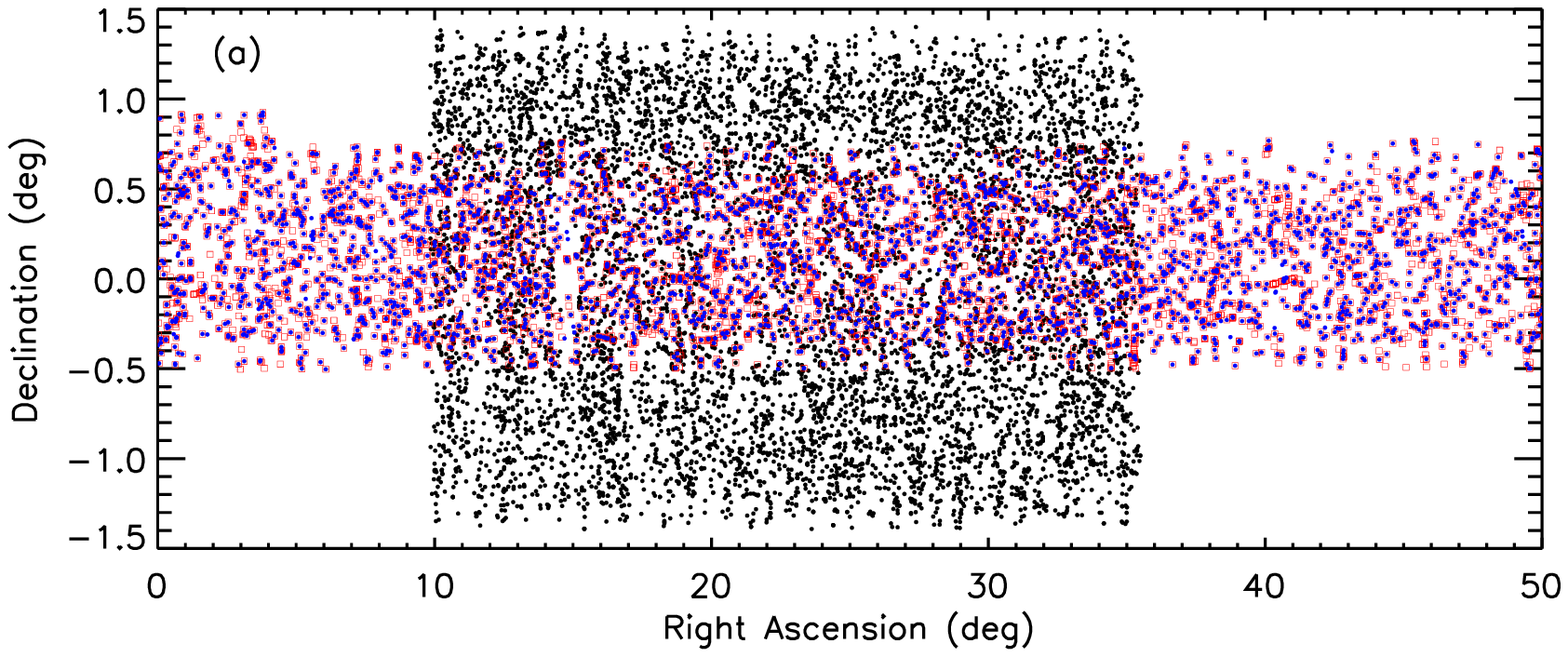}
\vspace{1pt}
\includegraphics[scale=0.75]{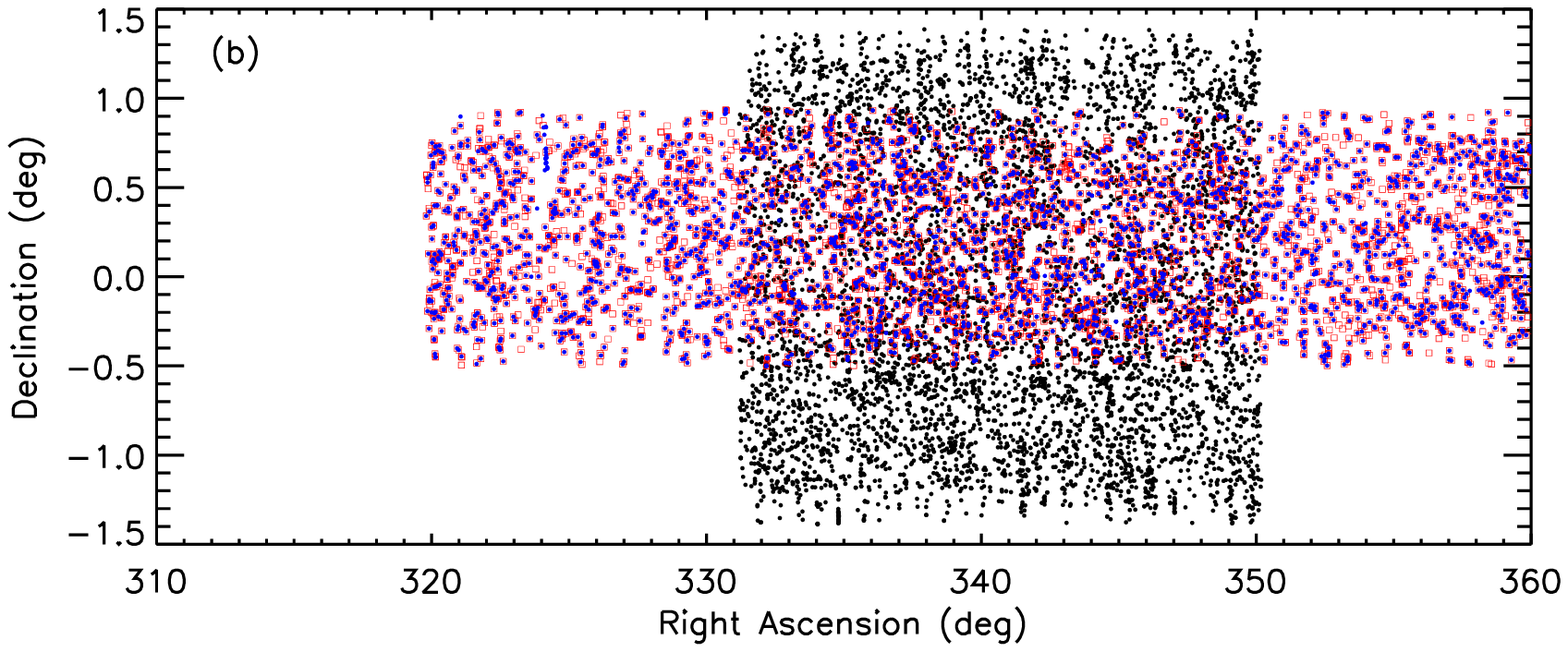}
\caption{Area covered by each of the three epochs.  Epoch I and II come from the FIRST survey and are shown as blue dots and red squares, respectively.  Epoch III is from a recent high--resolution VLA survey of Stripe 82 and is shown by the black dots. Note that the axes are not to scale.}
\label{fig:AreaCovered}
\end{figure*}
		
A number of papers, many using the first Allen Telescope Array (ATA) results, have recently been published on searches for transient sources \citep{Bower:2007p346, 2010ApJ...719...45C, 2011ApJ...739...76B, 2011ApJ...728L..14B}. Studies of long--term variability are harder to find.  One such study, by \citet{Becker:2010p157}, used three epochs of VLA observations to look for variable sources in the Galactic Plane.  Their 6 cm data covered 23 deg$^2$ of the Galactic Plane down to a limiting flux density of 1 mJy.  They found 39 variable sources, and they showed that these sources are more highly variable than extragalactic objects.  They also concluded that the variable fraction increases toward the Galactic center.


Out of the Galaxy, radio--loud active galactic nuclei (AGN) are by far the most common object in radio imaging surveys with flux thresholds above $\sim$1 mJy.  
Radio--loud AGN have been observed to vary on timescales from less than a day to years in all frequency regimes.
Some of the radio variability, particularly at low ($<$1 GHz) frequencies, may be extrinsic to the source.
For example, interstellar scintillation \citep[ISS;][]{Rickett:1986p564, 1987AJ.....93..589H} 
and extreme scattering events \citep[ESE;][]{Fiedler:1987p675, Romani:1987p324}
are intensity variations thought to be caused by the turbulent, ionized interstellar medium of our own galaxy.
However, ESEs are extremely rare, and ISS produces typical intensity variations of only a few percent \citep[e.g.,][]{2003AJ....126.1699L, 2008ApJ...689..108L}.
A majority of the observed extragalactic radio variability is therefore thought to be intrinsic to the source itself.

A popular theory for the instrinsic variability in extragalactic radio sources involves shock waves propogating along an adiabatic, conical, relativistic jet \citep{1985ApJ...298..114M}. 
In this model, the amplitude of variability should be larger (and the timescales shorter) for objects viewed close to the line of sight of the relativistic jet, such as BL Lac objects and flat spectrum quasars \citep{1978PhyS...17..265B, 1995PASP..107..803U}.
Extensive work has therefore been done on the extreme variability of bright blazars \citep[e.g.,][]{1992ApJ...386..473H, Aller:1999p601, 2004A&A...419..485C},
despite the fact that such sources make up only a small fraction of the total AGN population.
Other studies have targeted small groups of specifically--selected objects, such as the study by \citet{2001ASPC..224..265F} of 30 radio--quiet and radio--intermediate AGN, or they have concentrated on correlating shorter--term variability at multiple frequencies.  
There have been surprisingly few blind, long--term studies.  
 \citet{2003ApJ...590..192C} used multi--epoch VLA observations to study submillijansky variability in the Lockman Hole, producing just nine variable sources. 
\citet{deVries:2004p2565} used two VLA epochs of Faint Images of the Radio Sky at Twenty centimeters (FIRST) data to search for variability down to 2 mJy over 104 deg$^2$. 
More recently, \citet{2011ApJ...742...49T} used $\sim$55,000 FIRST survey snapshots to search for variable and transient objects on timescales of minutes to years.  This last survey produced 1,627 variable and transient sources down to millijansky levels, the largest sample to date.

In this paper, we use a new 1.4 GHz survey of the Sloan Digital Sky Survey (SDSS) Southern Equatorial Stripe \citep{2011AJ....142....3H} to further our understanding of long--term radio variability in faint extragalactic sources.  This survey of the Southern Equatorial Stripe, a.k.a.\ ``Stripe 82", covers 92 deg$^2$ to a typical rms of 52 $\mu$Jy, making it the widest survey to reach this depth.  These high--quality data overlap with the FIRST coverage of Stripe 82 examined by \citet{deVries:2004p2565} for variability, allowing us to extend their pilot study and add a third epoch of data with three times the angular resolution (1.8$^{\prime\prime}$). 

We begin by describing the radio data in Section \ref{data}.  The details of our sample selection are discussed in Section \ref{sample}.  We describe our results in Section \ref{results}, 
including sections on source strength dependence (Section \ref{flux}), 
fractional variability (Section \ref{varamp}), 
characteristic timescale (Section \ref{timescale}), 
optical counterparts (Section \ref{optcounterparts}), 
and morphology (Section \ref{morph}).  
We discuss the results in Section \ref{discussion}, 
and we end with our Conclusions in Section \ref{conclusions}.  
Where applicable we assume the standard $\Lambda$CDM cosmology of $H_0$ = 70 km s$^{-1}$ Mpc$^{-1}$, $\Omega_{\Lambda}$ = 0.7, and $\Omega_{M}$ = 0.3 \citep{2003ApJS..148..175S, 2007ApJS..170..377S}.

\begin{figure*}
\centering
\includegraphics[scale=0.72]{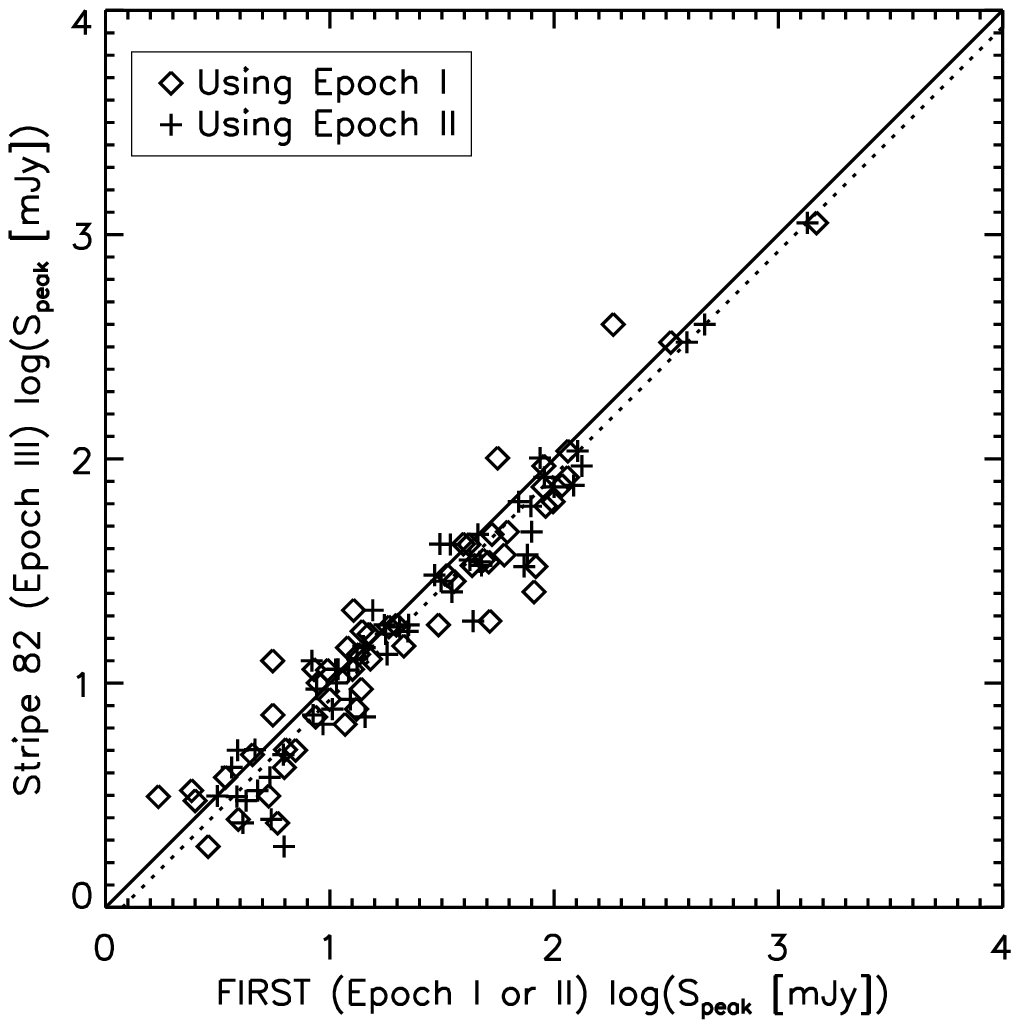}
\hfil
\includegraphics[scale=0.72]{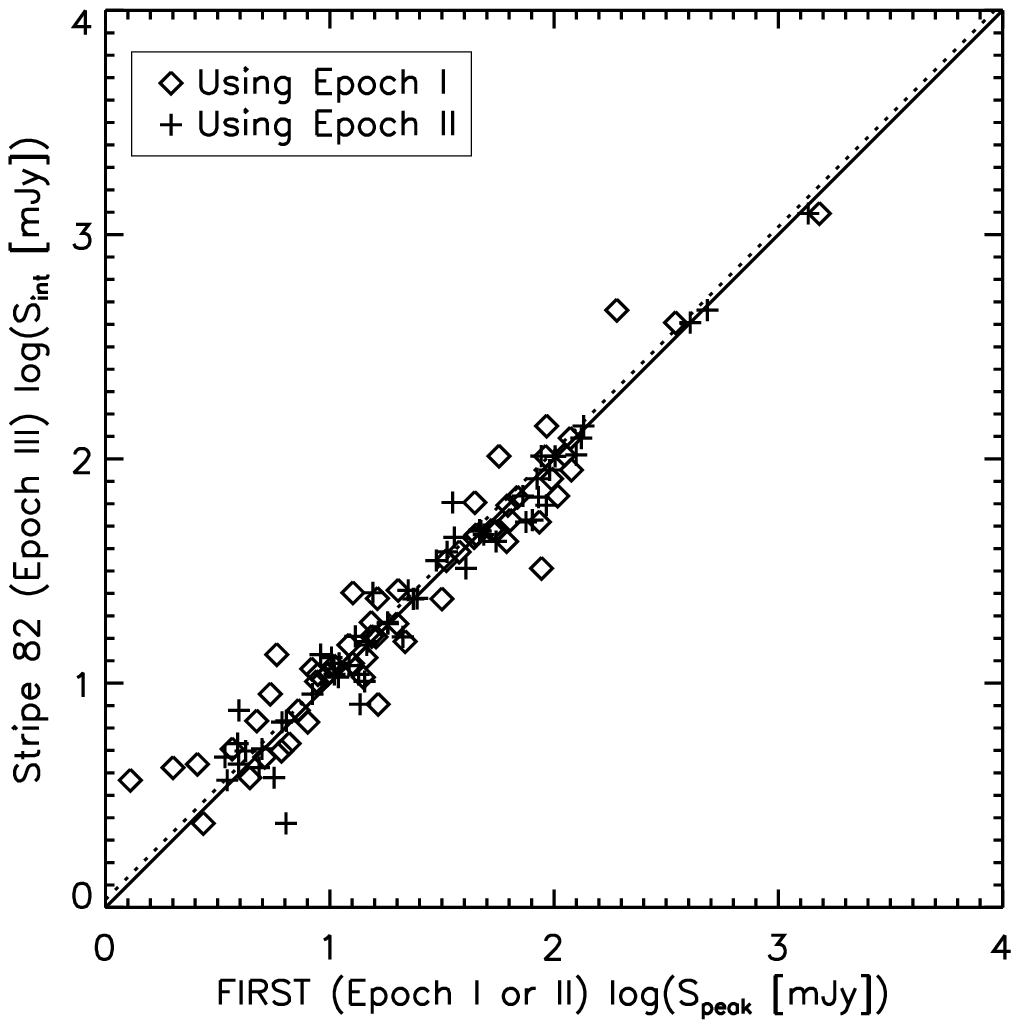}
\caption{Peak (left) and integrated (right) flux density from the Epoch III Stripe 82 catalog versus peak flux density from FIRST (Epoch I or II) for the variable sources previously identified by \citet{deVries:2004p2565}.  Variable sources that appeared in both the Epoch I and Epoch II catalogs are represented twice.  The typical error bars are much smaller than the plot symbols. The solid line indicates where the flux densities from the different epochs are equal.  The dotted line is a fit to the data.  The left panel shows that the Stripe 82 sources have peak flux densities that are 15\% fainter on average than the corresponding peak flux densities from FIRST. In the right panel, a fit to the data is again overplotted as a dotted line, but it lies basically on top of the solid line and is not as noticeable.  The bias indicated by the fit is 8\%, this time in the direction of higher Epoch III flux density values.}
\label{fig:FZDvariables_bias}
\end{figure*}


\section{Radio Data}
\label{data}

We used three different epochs of radio data to conduct our study of radio variability.  The first two epochs (1995 and 2002) come from the FIRST survey \citep{1995ApJ...450..559B}.  The FIRST survey was conducted with the Very Large Array (VLA) at 1.4 GHz and covered $>$9,000 deg$^2$ in the North and South Galactic caps.  The survey utilized the VLA's B--configuration, giving it a resolution of 5$^{\prime\prime}$.  The typical rms achieved was 0.15 mJy, and the catalog used a 1 mJy detection threshold.  

The Southern Equatorial Stripe was first observed by the FIRST survey in 1995.  The area was subsequently re--observed in 2002 for two reasons: as a quality control test of the FIRST survey, and to search for variability.  The area that was re--observed is smaller than the initial area observed, covering 104.3 deg$^2$  
and containing 9,086 radio sources.  
We will refer to the 1995 and 2002 epochs as Epoch I and Epoch II, respectively.  

The third epoch of radio data comes from a recent high--resolution VLA survey of SDSS Stripe 82 \citep{2011AJ....142....3H}.  This 1.4 GHz survey covered 92 deg$^2$ to a typical rms of 52 $\mu$Jy, making it the largest 1.4 GHz survey to reach such a depth.  The majority of the data were taken in the A--configuration, with supplemental B--configuration data obtained on every field to increase sensitivity to extended structure.  The resolution achieved was 1.8$^{\prime\prime}$, or roughly three times better than FIRST.  These data constitute Epoch III and may sometimes be referred to as the `Stripe 82' radio data.

The area covered by Epoch III does not entirely overlap the area covered in Epochs I and II.  Figure \ref{fig:AreaCovered} 
shows the area covered by the individual epochs.  The SDSS Stripe 82 data (Epoch III) avoided the region around RA $=$ 0, as planned infrared observations would be compromised by zodiacal light in that part of the sky.  The total area observed by all three epochs is 60 deg$^2$.  

\citet{deVries:2004p2565} previously searched for variables in Stripe 82 using the data from Epochs I and II.
In the course of their analysis, they determined that the flux densities of the Epoch II data required a small correction factor consisting of a 90 $\mu$Jy zero--point offset and a 1.16\% sensitivity correction, well within the estimated $\sim$5\% systematic uncertainty in the flux density scale.
In the following analysis, we have used the Epoch II data with this correction applied.



\section{Sample Selection}
\label{sample}
\subsection{Variable Sources}
This work builds on the work that \citet{deVries:2004p2565} previously did to search for variability between the two FIRST epochs (I and II).
Of the 128 variable sources found by \citet{deVries:2004p2565} between Epochs I and II, 58 fall in the area covered by the Epoch III observations.  
Before searching for new variables with the Epoch III data, we first matched these known variable sources to the Epoch III data to look for systematic effects between the epochs.

Using a 3$^{\prime\prime}$ matching radius, we matched these sources to the Epoch III data.  We chose this matching radius as it is approximately half of the FWHM of the beam from the two earlier FIRST epochs.  We found that 57 of the 58 sources were recovered in the Epoch III catalog, with the single source that was not recovered falling just below the detection threshold for the Epoch III catalog.  We obtained the peak and integrated flux density of this source by using the task JMFIT in AIPS.

Figure~\ref{fig:FZDvariables_bias}a shows the Epoch I/II peak flux density of the previously known FIRST variables plotted against the peak flux density from the new Epoch III observations.  
We plot peak flux density here instead of integrated flux density as variable sources are expected to be point--like, and the peak flux density is a better approximation of the actual flux density for unresolved sources.  
Each variable source is represented on this plot twice: once with its Epoch I flux density, and once with its Epoch II flux density.  
The solid line has a slope of log(x) $=$ log(y) and should roughly bisect the sources if they are getting brighter or fainter at random.  
What we see, however, is a trend toward fainter Epoch III flux densities.  
This is especially suspect, as the Epoch III Stripe 82 observations are the higher--resolution observations.  
To quantify this bias, we fit a line to the data, fixing the slope to 1.0, but allowing the intercept to vary.  
We found that the Epoch III observations have peak flux densities that are 15\% ($\pm$2\%) fainter on average.  
The measured bias indicates that these sources may be resolved by the higher resolution Epoch III data, necessitating the use of integrated flux density for Epoch III.  
Figure~\ref{fig:FZDvariables_bias}b therefore shows the same sources, but with integrated flux density instead of peak flux density for Epoch III.    
The bias we calculate (8\% $\pm$ 2\%) is now in the direction of slightly larger Epoch III flux densities,
and our estimate remains the same if we restrict the sample to those sources that are also point sources in Epoch III.
This offset is likely indicative of the systematic bias due to, for example, errors in absolute flux calibration.  
Due to the improvement in the correlation seen in Figure~\ref{fig:FZDvariables_bias}b, we therefore find it necessary to use integrated flux density when comparing Epochs I/II with Epoch III. 

\begin{figure}
\centering
\includegraphics[scale=0.74]{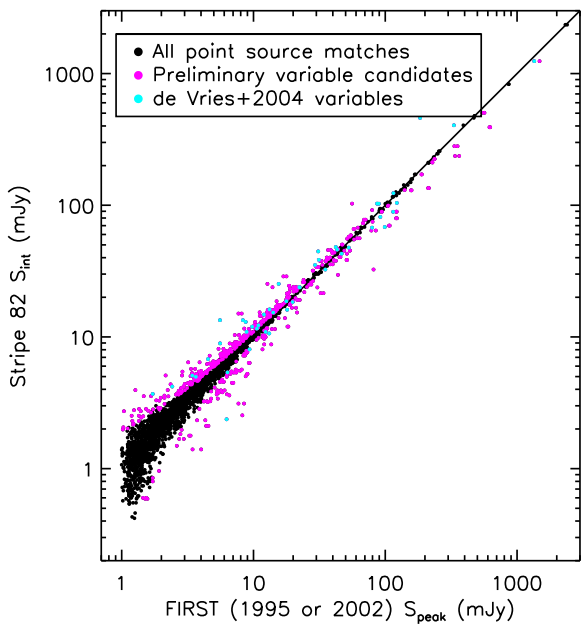}
\caption{Epoch III (Stripe 82) integrated flux density versus Epoch I or II FIRST peak flux density for all sources which satisfy our point source criterion and have a match between Epoch III and at least one of the earlier epochs. Sources in both FIRST epochs are plotted twice -- once with each FIRST flux -- and may be identified as variables with respect to Epoch III in one or both cases. Sources that exceed our preliminary variability threshold (Equation 1) are shown in pink, while sources previously--identified as variable by \citet{deVries:2004p2565} are shown in cyan. }
\label{fig:variables_int}
\end{figure}

We then used the high--resolution Stripe 82 data to search for new variable sources. 
We matched the Epoch III (Stripe 82) source catalog to both the Epoch I and Epoch II FIRST catalogs, again using a matching radius of 3$^{\prime\prime}$. We required the sources to be point sources in the Stripe 82 data, where we defined a ``point source" as anything with a ratio of peak--to--integrated flux density S$_{\rm III,pk}$/S$_{\rm III,int}$ $>$ 0.7.  
The exclusion of extended sources from a variability search is justified in that extended sources are not expected to be variable on such short timescales (less than $14/[1+z]$ years).
Upon further investigation, we found that this point--source criterion results in the elimination of some of the previously--known variables \citep[from][]{deVries:2004p2565} 
from our candidate list due to the higher--resolution of the new data.
It is therefore possible that some variable sources are missed by this cut. 
Indeed, we found that adjusting the value of this cutoff can have a large effect on the number of variable sources in our sample.  
However, for the purposes of this initial variability census, we argue that reliability is more important than completeness, and we have required all variable sources to meet this criterion.  

We defined a preliminary variable sample consisting of all point sources which satisfy the criterion:

\begin{equation}
\Delta {\rm S} > 5\times \left ( {\rm \sigma}_{\rm I,II}^{2} + {\rm \sigma}_{\rm III}^{2} \right )^{1/2}
\end{equation}
\label{eqn:variability}

\noindent where $\Delta {\rm S}$ is the flux density difference between the two epochs being compared,
$\sigma_{\rm I,II}$ is the rms at the position of the source in either Epoch I or II (depending on which is being matched), and 
$\sigma_{\rm III}$ is the error on the integrated Stripe 82 flux density. 
Note that we also required a variability amplitude of $>$5\% to account for uncertainties in absolute flux calibration.
For the FIRST epochs I and II, the rms is a good estimate of the error on the peak flux density,  which we used as the best estimate of the true flux density.
For Epoch III, on the other hand, we used the integrated flux density (as discussed earlier in this section), 
and since the uncertainty on the integrated flux density is greater than the rms listed in the catalog at the source position, we used the former as the flux density uncertainty.
These errors were calculated by the automated source--finder HAPPY during the creation of the Stripe 82 catalog \citep[see][]{2011AJ....142....3H}. 
For all sources, the Poisson noise is negligable.

\begin{figure}
\centering
\includegraphics[scale=0.5]{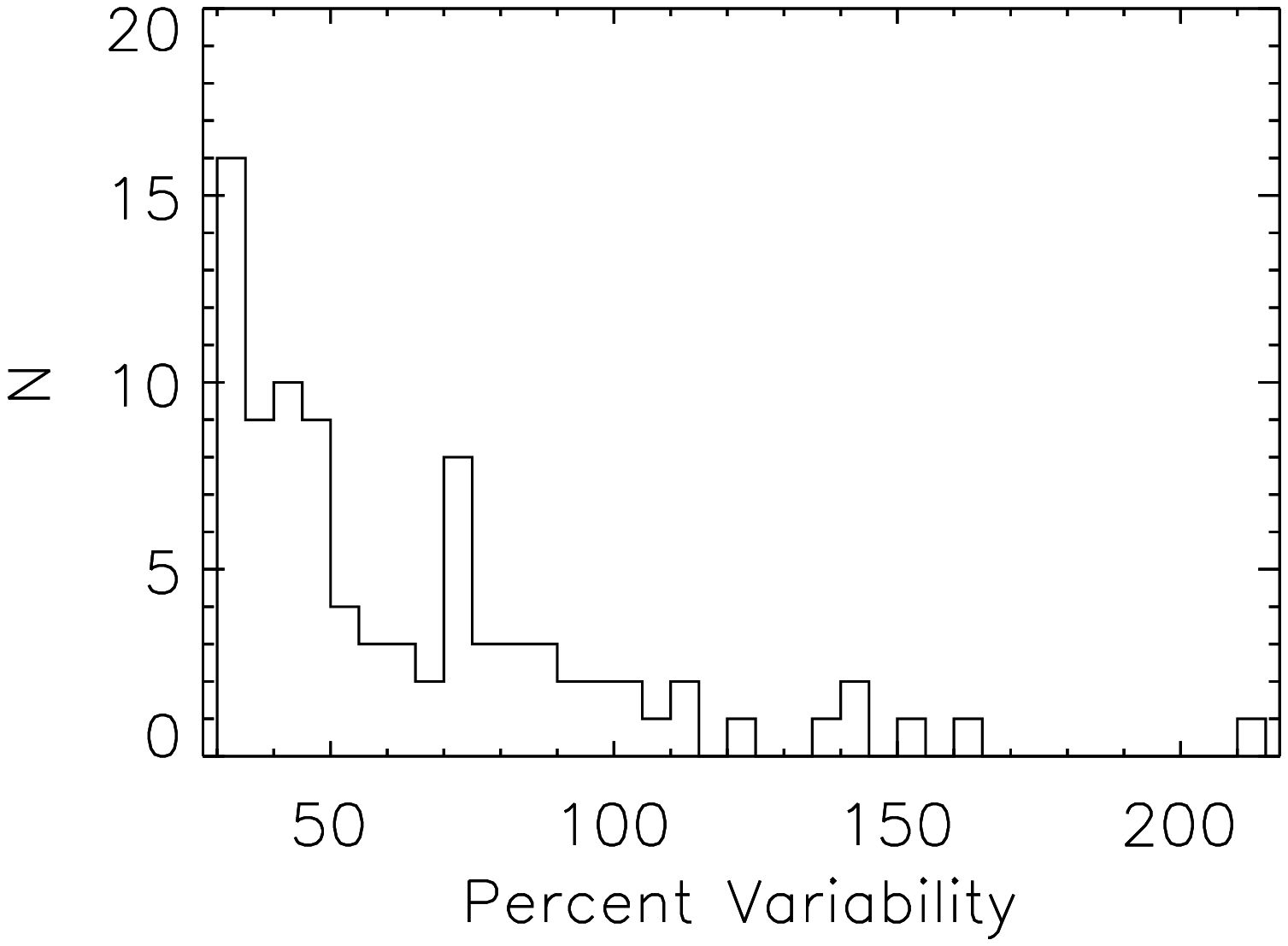}
\vfil
\includegraphics[scale=0.5]{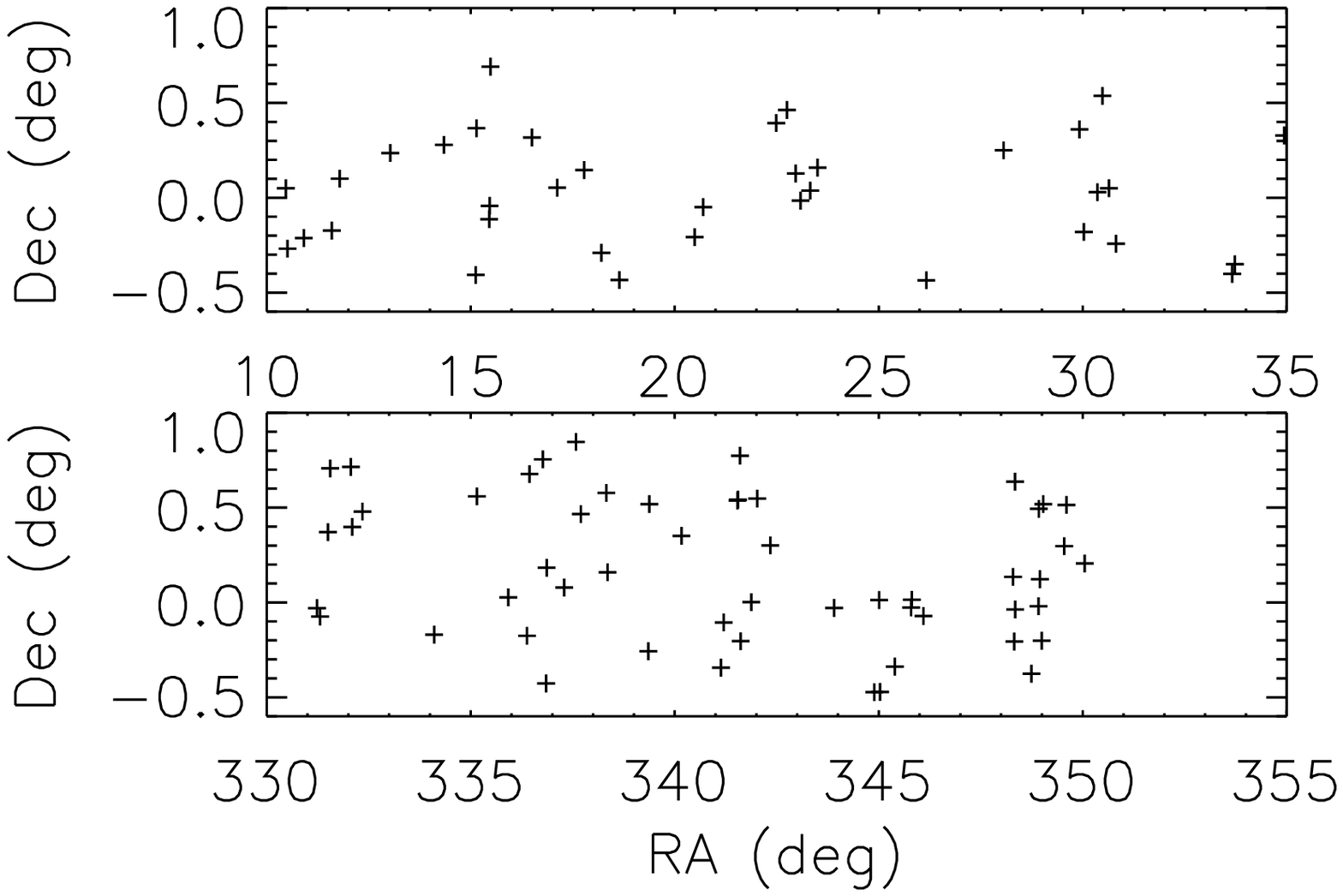}
\vfil
\includegraphics[scale=0.5]{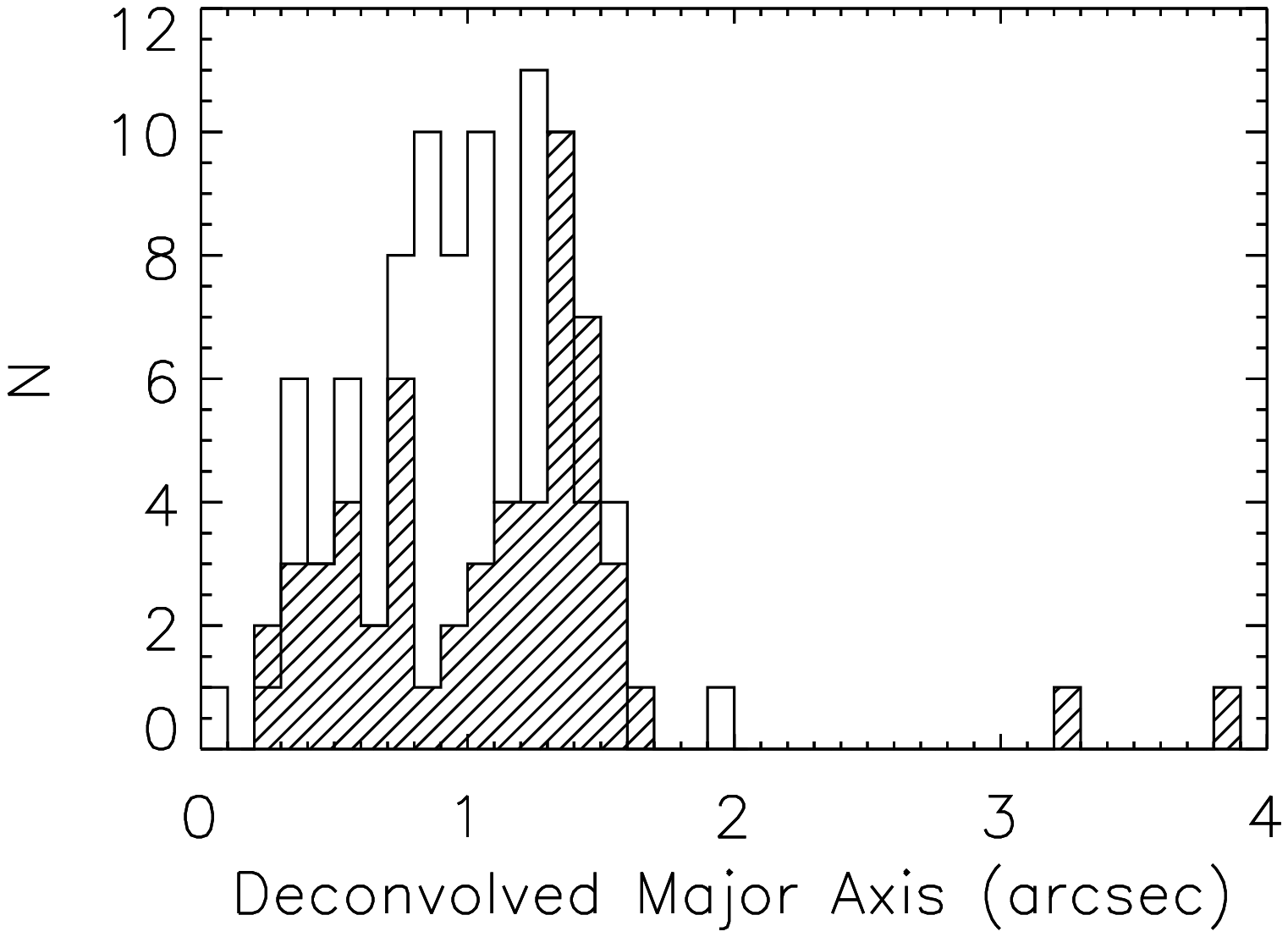}
\caption{\textit{Top:} A histogram showing the amplitude of the variability (as a percent of the source strength) for the final variable sample.  Due to the possible presence of systematic effects such as absolute flux calibration errors, we only consider sources which vary by more than 30\%. \textit{Middle:} The distribution on the sky of the final sample. The area with RA$>$350$^{\circ}$ is not covered by this study. \textit{Bottom:} Distribution of the deconvolved major axis values for the previously--identified variable sources from \citet{deVries:2004p2565} (filled histogram) and the variable sources identified with the addition of Epoch III data (plain histogram).
}
\label{fig:variables_props}
\end{figure}

We identified 1,436 distinct point sources (Figure~\ref{fig:variables_int}) with a match between the Epoch III catalog and at least one of the earlier epochs. 
Of these point source matches, 258 satisfied the preliminary variability criterion defined in Equation~1.  
These variable source candidates (shown in pink) lie, by definition, on the outer envelopes of the distribution.  
One of the more obvious features of the plot is that the distribution of matches is asymmetric; there are actually \textit{more} sources that appear brighter in Epoch III.  
This is an effect of the Malmquist bias, and it was discussed in this context (Stripe 82 versus FIRST survey) in the Stripe 82 VLA survey paper \citep{2011AJ....142....3H}.  
For a source that is resolved in the Stripe 82 data, Stripe 82 is not as sensitive as FIRST, which can cause the peak flux density of the source to fall below the detection threshold.  
It will appear in the Stripe 82 catalog only if a noise fluctuation, or variability, causes the apparent peak flux to be higher.  
In that case, the source will also be brighter than the FIRST source.  
The bias seen in Figure \ref{fig:variables_int} is therefore simply an artifact of the difference in resolution between the epochs. 

A second effect is visible at the bright end of the distribution.
There it appears that bright sources tend to be fainter in the Stripe 82 epoch, even despite the use of integrated Stripe 82 flux density. 
Due to the clear presence of resolution effects, therefore, 
we have conservatively defined as variable only those sources which have a higher flux density in the high--resolution (Epoch III) data.   

Finally, flux calibration plays an extremely important role in any study concerned with variability.  If the absolute flux calibration differs between year to year or field to field, this could affect how many sources are flagged as variable.  To estimate the effect of a flux calibration error, we tried increasing and decreasing the Stripe 82 integrated and peak flux densities by small amounts.  We found that if we altered the Stripe 82 fluxes by even 5\% (the value assumed above for the calibration uncertainty), the number of variable candidates changed dramatically (by $\sim$50\%).  If the absolute flux calibration is slightly different on certain fields, this might therefore cause a non--uniform distribution of the variable sources.  We cannot attempt to correct for this effect by adjusting the flux density ratios to a median value of unity since the the Malmquist effect and remaining resolution bias are uncertain.  
Because of this uncertainty, therefore, we imposed one final, additional constraint: we consider as variable only those sources with a flux density variation greater than 30\%.  
Such large variations are less sensitive to smaller systematic effects like differences in absolute flux calibration.


With these additional criteria, we defined a final variable sample of 89 sources.
These 89 sources will be the variable sample we refer to for the remainder of the paper.  
Of these sources, 14 were previously identified by \citet{deVries:2004p2565}, and 75 are newly--identified.  
Note that, because we only considered sources that are brighter in the high--resolution data, we do not find it necessary to disallow multiple Stripe 82 sources within the 3$^{\prime\prime}$ radius, which could be due to a FIRST source being resolved into multiple components.  If a FIRST source were to be resolved in the Stripe 82 data, the flux density of the central component would appear fainter, not brighter.

Some basic properties of the final variable sample are shown in Figure \ref{fig:variables_props}.
A histogram of percent variability displayed by the variable sources is shown in Figure \ref{fig:variables_props} (top),
where all sources vary by $>$30\% as outlined above.
The distribution on the sky of these sources
is shown in Figure \ref{fig:variables_props} (middle).
This plot confirms that the variable sources are not densely clustered in certain areas of the stripe, a result which would imply errors in photometric calibration rather than true variability.
Figure \ref{fig:variables_props} (bottom) shows the distribution of deconvolved major axis values. 
The open histogram shows the final variable sample, while the filled histogram shows the newly--derived values for the previously--identified variables \citep{deVries:2004p2565} using the higher resolution Epoch III data.
The distribution for our final variable sample extends up to major axis values of 2$^{\prime\prime}$.  
Aside from the two outliers with Maj $>$ 3$^{\prime\prime}$, the previously--identified variables all have values of Maj $<$ 1.7$^{\prime\prime}$.  
We can therefore see that if we applied our point source criterion to the \citet{deVries:2004p2565} sample, we would be 96\% complete, as 56 out of their 58 sources (covered here) have major axis values below 2$^{\prime\prime}$.  

In summary, our final variable sample consists of sources which have a match (within 3$^{\prime\prime}$) between the Stripe 82 data (Epoch III) and one or both FIRST epochs, and which meet the following four criteria:
\vspace{3mm}

\indent (1) Display flux density differences greater than 5$\sigma$ (Equation 1)\\
\indent (2) Point source in the high--resolution Epoch III data (i.e.\ S$_{\rm III,pk}$/S$_{\rm III,int}$ $<$ 0.7)\\
\indent (3) Brighter in the high--resolution Epoch III data (i.e.\ S$_{\rm III}$ $>$ S$_{\rm I,II}$)\\
\indent (4) Variability amplitude $>$ 30\% 

\vspace{3mm}
The application of these criteria results in 89 sources, and these source make up our final variable sample.

\begin{figure}
\centering
\includegraphics[scale=0.7]{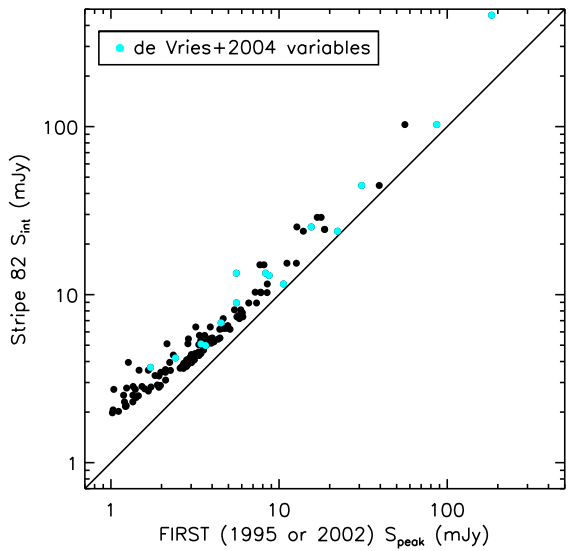}
\caption{Epoch III (Stripe 82) integrated flux density versus Epoch I or II FIRST peak flux density for our final variable sample.  Sources from the sample that were previously identified by \citet{deVries:2004p2565} are shown in cyan.}
\label{fig:variables_bright}
\end{figure}

\subsection{Transient Phenomena}
\label{transients}
Transient phenomena, although rare, can reveal interesting and new astrophysical sources when discovered.  We searched for transient radio sources in our data in much the same way that we searched for variability.  Due to the resolution bias, we only searched for sources that were undetected in the FIRST data, and not the other way around, since FIRST sources may simply be resolved out of the Stripe 82 data.  We considered only the Stripe 82 sources that satisfied our point source criterion, and for each undetected source, we retrieved the FIRST rms in each epoch at that location from the respective coverage maps.   We required the source to satisfy our variability criterion, assuming an upper limit for the source of 5$\times \sigma_{\rm FIRST}$ $+$ 0.25 mJy to account for the FIRST CLEAN bias \citep{1995ApJ...450..559B}.  We also required sources to be absent from both FIRST epochs, as \citet{deVries:2004p2565} reported a systematic effect which caused a large number of sources to appear only in one FIRST catalog or the other at faint flux densities.  This left us with 14 initial transient source candidates.  

Upon close inspection of each of the candidates, none appeared to be real transient sources.  The majority were sources that had been resolved in the Epoch III data such that the position of the FIRST component shifted by more than 3$^{\prime\prime}$, or another component appeared 3$^{\prime\prime}$ from the original component.  This was not the case for two of the sources, which were clearly isolated point sources in the Stripe 82 images.  However, we detected FIRST counterparts to both of these sources in the combined (Epoch I$+$II) FIRST catalog. 
Finally, one of the sources was a strong sidelobe ($S_{\rm int}$ = 4.3 mJy) of a bright source roughly 2$^{\prime}$ away.  

Therefore, we did not find any evidence for transient phenomena in the Stripe 82 data.  
We note that this result is not particularly surprising, particularly since both the FIRST and Stripe 82 catalogs are based on coadded images using individual observations spread over days to months. 
Transients that appear on shorter timescales would therefore appear fainter in the coadded data, rendering them harder to detect.

\begin{figure}
\centering
\includegraphics[scale=0.5]{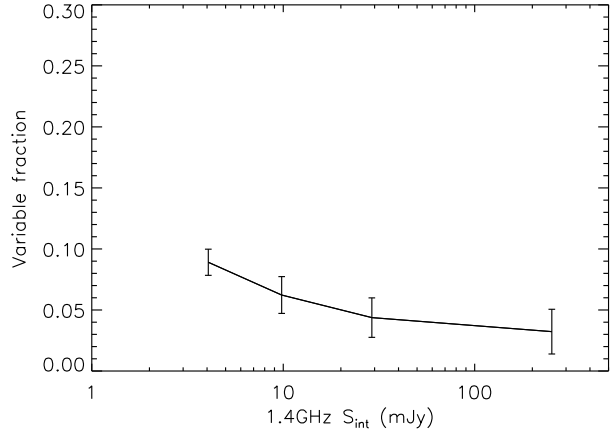}
\caption{Variable fraction of radio sources as a function of flux density.  The median flux density for each flux density bin is plotted. }
\label{fig:varfract}
\end{figure}





\section{Results}
\label{results}
Figure \ref{fig:variables_bright} is the main figure that characterizes the newly--defined variable sample.  
It shows the final variable sample, with those that were previously identified by \citet{deVries:2004p2565} overplotted in cyan.  
We find that 14 of the 58 previously--identified variables satisfy our variability criterion.  
Note that this means they have increased in brightness, as we are limiting our variable sample to sources that are brighter in the high resolution epoch.  
Of the remaining de Vries et al. variable sources, two sources did not survive our point source requirement and the remaining sources were either not found to be variable, or varied significantly in flux density but were fainter in Stripe 82, and thus ambiguous.  
Assuming that the number of sources that get brighter is roughly equal to the number of sources that fade with time, we estimate that the true number of variable sources (based on our other criteria) is double the number we find here, or approximately 28 sources of the 58 previously identified.  

Table \ref{tab:EpochIIIvariables} lists the 89 variable sources discovered by matching the two FIRST epochs to the Epoch III data.  Variable sources that were previously discovered by \citet{deVries:2004p2565} are indicated with an asterisk before the RA.  Values for the peak and integrated flux density and rms for each of the three epochs are listed, as well as the deconvolved major axis value measured from the Epoch III data, the variability amplitude $f_{\rm var}$ (Section \ref{varamp}), and the morphological classification in FIRST and Stripe 82 (Section \ref{morph}).  Values of 0.0 indicate that the source was not detected in that epoch.  Note that strong sources significantly affect the local noise, causing the rms listed to increase dramatically from the median survey value of 52 $\mu$Jy.  Assuming that an equal number of sources increase and decrease in flux density, we would expect approximately 178 variable sources total in the area, or $\sim$3 variables deg$^{-2}$.

Note that we used a 5$\sigma$ variability threshold instead of the 4$\sigma$ threshold used by \citet{deVries:2004p2565}.  (They also used a slightly different equation -- see \citealt{deVries:2004p2565} for details.)  However, Epoch III has a lower typical rms, and is therefore sensitive to lower levels of variability.  This explains why a small number of the newly detected variable sources were actually slightly \textit{more} variable between Epoch I and II than they were between Epochs I/II and III, and yet were not identified by \citet{deVries:2004p2565} as variable sources.

Table \ref{tab:FZDvariables} lists all of the 58 sources that were part of the \citet{deVries:2004p2565} sample and occur in the area covered by the new Epoch III data.  The columns are the same as Table \ref{tab:EpochIIIvariables}.  Sources that meet our variability criteria for Epoch III (including a brighter flux density in the new epoch) are again indicated with an asterisk before the RA.  As the sources in this sample tend to be much brighter, in general, than the sources selected with the Epoch III data, the local rms values are also higher.

\begin{figure}
\centering
\includegraphics[scale=0.5]{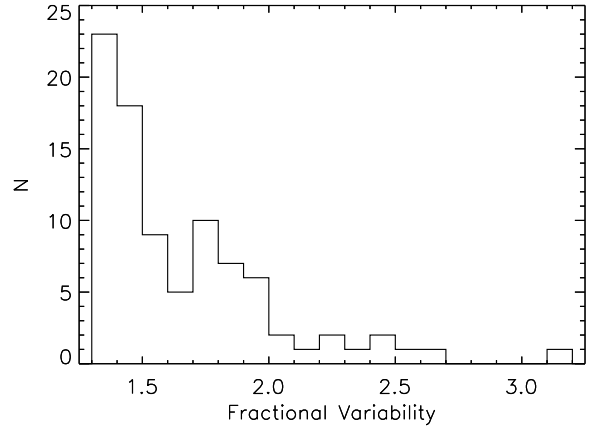}
\caption{Histogram of fractional variability of the variable sources, defined as the maximum integrated flux density over the minimum integrated flux density.  Note that since we require sources to vary by at least 30\%, nothing has a fractional variability of less than 1.3.}
\label{fig:variables_fvar}
\end{figure}

\subsection{Dependence on Source Strength}
\label{flux}
To investigate whether the variable fraction varies with flux density, we split the variable sample into four flux density bins. 
The median flux values of the bins ranged from 4 mJy on the faint end to 250 mJy on the bright end.  
The variable fraction for each bin is plotted against the median flux density in Figure \ref{fig:varfract}. 
We find that the variable fraction remains essentially constant as a function of flux density, with a possible increase in the fraction of variable sources at fainter flux densities (though we caution that a slightly different binning can remove this subtle effect, and it may therefore not be significant).
This is despite the fact that the data are inherently more sensitive to variability in very bright sources, while for faint sources it takes a large change in amplitude to achieve the 5$\sigma$ threshold.  
Our requirement of 30\% variability makes the sample more consistent, but the variable fraction is very sensitive to the exact cut applied. 
Therefore, it appears that 
faint radio sources are just as variable as bright radio sources, and possibly more so.

\subsection{Fractional Variability}            
\label{varamp}
A paper by \citet{Becker:2010p157} on variable sources in the Galactic plane found that Galactic sources were more variable than extragalactic sources.  Their Galactic sample came from three epochs of VLA data on the Galactic Plane taken roughly 15 years apart, and they compared this against the extragalactic sample of \citet{deVries:2004p2565}.  To see if this still holds true for our more robust sample of extragalactic variables, we calculated the fractional variability for the sources 
in our variable sample, where fractional variability is defined as:
\begin{equation}
f_{var} = \frac{S_{\rm max}}{S_{\rm min}}
\end{equation}
where S $=$ S$_{\rm pk}$ for the FIRST epochs I/II and S $=$ S$_{\rm int}$ for Epoch III.

Figure \ref{fig:variables_fvar} shows the distribution of values for the fractional variability.  
Note that since we required a variability amplitude of $\geq$ 30\% to be in the variable sample, no sources have a fractional variability less than 1.3.  
Our values are compared to the extragalactic sample of \citet{deVries:2004p2565} and the galactic sample of \citet{Becker:2010p157} in Table \ref{tab:fvar}.  
The bins are as they were defined in \citet{Becker:2010p157}. 
The lowest bin is empty for our sample and the second--lowest bin only includes sources with fractional variabilities down to 1.3 due to our variability amplitude requirement.  
Nevertheless, there is a clear difference in the distribution of fractional variability between our sample and the Galactic sample.  
Ignoring the lowest bin, almost half (46\%) of our extragalactic sources have a fractional variability $f _{var}< 1.5$, while only 13\% of Galactic sources fall in the same range.  
Meanwhile, only 1\% of extragalactic sources have a fractional variability  $f_{var}$ $>$ 3 (corresponding to a variability amplitude of 200\%), compared to a full 44\% of Galactic sources.  
The difference between the samples is significant at the $>$5.5$\sigma$ level based on a $\chi^2$ contingency test.  
Therefore, we confirm that Galactic sources appear to be more highly variable than extragalactic sources.  

\begin{figure}
\centering
\includegraphics[scale=0.5]{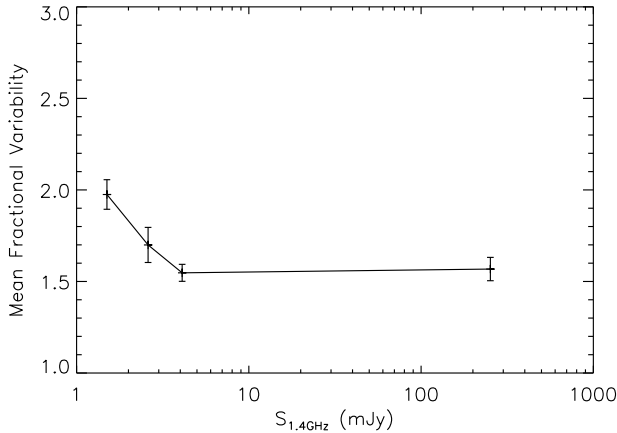}
\caption{Mean fractional variability as a function of flux density.  We took the average flux density for each individual variable source, and the plotted points show the average of those individual values within a bin. }
\label{fig:fvar_vsFlux}
\end{figure}

We show the flux dependence of the fractional variability in Figure~\ref{fig:fvar_vsFlux}, where we plot mean fractional variability as a function of flux density bin.
At low flux densities, where most of the variable sources lie, there appears to be a steep rise in fractional variability.
Since our variable sample has roughly the same flux density distribution as the Galactic sample, this trend cannot explain the very different fractional variability distributions of the two samples. 
Within our sample, it is likely a selection effect, since (as discussed in Section~\ref{flux}) fainter sources must exhibit a larger (fractional) change in amplitude in order to exceed the 5$\sigma$ threshold in the first place.

\subsection{Characteristic Timescale}
\label{timescale}
We can attempt to infer something about the characteristic timescale of AGN variability by looking at the number of variables discovered between the three epochs.  Splitting our results up by epoch, we find 55 variable sources between Epochs II and III (a 7 year baseline) and 74 variable sources between Epochs I and III (a 14 year baseline).  These numbers do not add up to the 89 sources in our final sample because some sources registered as variable in both searches.  

However, we cannot directly compare these numbers to the 58 variables found by  \citet{deVries:2004p2565} between Epochs I and II since we used different variability criteria and the new data reach a lower rms sensitivity.  
For a fairer comparison, we redid the search for variables between the FIRST epochs (I and II) using our Equation 1 and requiring $>$22\% variability.
(Note that the requirement for comparisons with Epoch III was 30\%, but the small, 8\% bias in the direction of higher Epoch III fluxes (Figure~\ref{fig:FZDvariables_bias}) means that the correct value to use here is really $30\%-8\% = 22\%$.)
To address the different rms sensitivities reached, we also redid our variability search between Epochs I/II and Epoch III, substituting the FIRST rms of 0.15 mJy for the Epoch III rms  
and scaling it up by $\sqrt{\rm N_{beams}}$ since we are using the integrated Stripe 82 flux. 
Finally, we multiplied these results by a factor of two, assuming that the same number of sources fade as get brighter.

With these changes, we find 64 variable sources between Epochs I and II, 80 variables between Epochs II and III, and 104 variables between Epochs I and III.  
The timescale between Epochs I--II and II--III is seven years in both cases, so we would expect to find roughly the same number of variables each time.  
The numbers we get are consistent within 25\%, with the remaining difference likely explained by the intricacies of selecting a consistent sample across datasets with different angular resolutions and sensitivity limits.
The number of variables discovered between Epochs II and III (80) and Epochs I and III (104) are more reliably compared, as they both involve a FIRST epoch matched to the Stripe 82 epoch, and the flux calibration of the two FIRST epochs has been verified against one another (requiring only a small correction factor -- see Section~2).  
Of the 104 sources discovered using the 14--year time span, 62 were already detected in the 7--year span, and 42 are newly detected.  
It thus appears that lengthening the observing timescale by 7 years only produced $<$50\% more variables.  
While we do not have enough information to be quantitative, it thus appears that the average characteristic timescale for variability for these sources may be less than 14 years.

\subsection{Optical Counterparts}
\label{optcounterparts}



We now turn to the optical properties of the sample.  To investigate the nature of the variable sources, we matched our sample to three different optical catalogs: the general catalog of all sources from the coadded imaging data in Stripe 82, a catalog of spectroscopically--confirmed quasars from SDSS, and a catalog of photometrically--selected quasars from SDSS.  In order to interpret the results, we created a control sample of non--variable sources with which to compare.  As with the variable source sample, we required the ratio of peak to integrated flux density $S_{\rm peak}$ / $S_{\rm int}$ $\geq$ 0.7.  Whereas for the variable sample we required the change in amplitude to be $\geq$ 5$\sigma$, for the non--variable sample we required the change in amplitude to be $\leq$ 0.5$\sigma$.  Since variability is really a spectrum, this restriction ensures that we are comparing to the least variable objects possible.  

We further required that the control sample have the same flux density distribution as the variable sample.  
To do this, we binned the integrated flux density of the variable sample into logarithmic bins of 0.3 dex distributed from 1 mJy to 1000 mJy,
and we required the control sample to follow the same distribution.
We 
used a random number generator to select 
the non--variable sources 
from each bin to go into the control sample.

\begin{figure}
\centering
\includegraphics[scale=0.5]{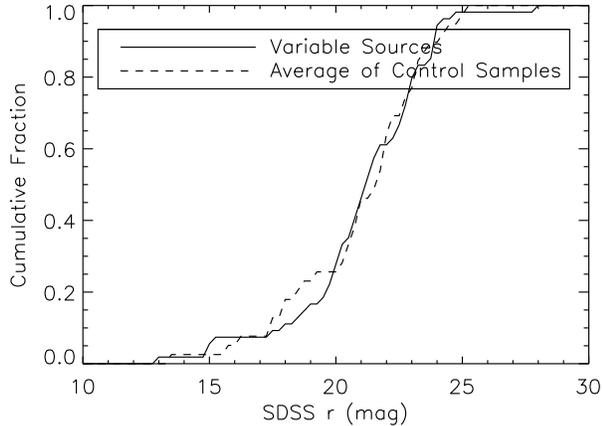}
\caption{Cumulative distribution function of SDSS r--band magnitude for the variable sources (solid line) and non--variable control sample (dashed line).}
\label{fig:SDSSrmag}
\end{figure}

If the variable sources are brighter than the control sample sources in the optical, this could bias the results, causing a higher overall match rate.  
In addition, brighter sources are easier to classify into galaxy/stellar sources, with fainter sources more likely to be classified as galaxies.  
To see if the magnitude distributions for the variable/non--variable samples are similar, we retrieved SDSS r--band magnitudes for the sources with optical matches in the variable and non--variable samples.  The cumulative distribution function of these magnitudes is shown in Figure \ref{fig:SDSSrmag} for both the variable sources (solid line) and non--variable control sample (dashed line).  
We used the Kolmogorov--Smirnov test to determine if the two magnitude distributions were likely to be drawn from the same underlying distribution.  The resulting KS statistic is 0.12 with a probability of 0.85 (where small values of the probability indicate that the cumulative distribution functions are significantly different).  We conclude that the difference between the magnitude distributions is most likely not significant and should not significantly influence star/galaxy classification or the overall match rate.  

We first matched the variable and control samples to the optical Stripe 82 catalog, created from combining the multiple epochs of SDSS imaging data of Stripe 82 into a single catalog that extends two magnitudes deeper than the main SDSS imaging area \citep{2011arXiv1111.6619A}.  After trimming this deep catalog to the area covered by the variable source sample, there are still over four million sources in the resulting optical catalog.   By matching to these sources, we find that the overall match rate for the variable source sample is 60\% $\pm$ 5\%, while the match rate for the control sample is 44\% $\pm$ 5\%.  Since it has been shown that the astrometry of the Stripe 82 radio catalog is well--tied to the SDSS \citep{2011AJ....142....3H}, we used a matching radius of 1$^{\prime\prime}$ to minimize the number of false (coincidence) matches.  To determine the rate of coincidence matches, we generated a fake catalog by shifting the radio source positions by 1$^{\prime}$ in four different directions and repeating the matching, averaging the results.  The false match rate we determined is 2\%, meaning that the vast majority of the matches reported above are likely to be real.  

\begin{figure}
\centering
\includegraphics[scale=0.5]{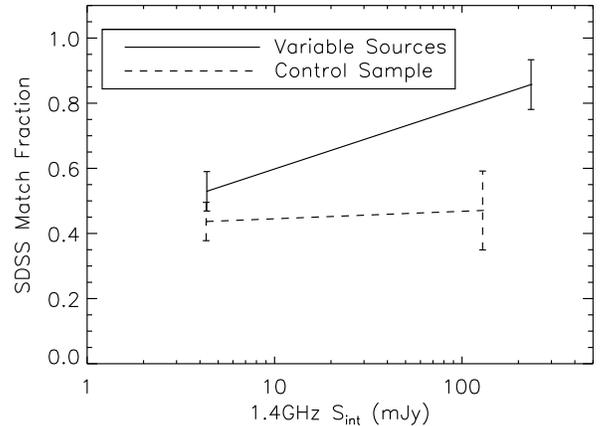}
\caption{Fraction of sources that match to the SDSS Stripe 82 optical catalog as a function of integrated flux density.  Variable sources are shown by the solid line, and the non--variable control sample are shown by the dashed line.  The data were split into two bins, and the data points indicate the centers of those bins.}
\label{fig:SDSSmatchrate_variables}
\end{figure}

To see if the results depend on flux density, we plot SDSS match fraction as a function of flux density in Figure \ref{fig:SDSSmatchrate_variables}.  
The variable sources show 
an increase in match fraction for brighter sources, while the non--variable sources show a matched fraction that is independent of radio flux density.  
While the matched fraction for the two samples appears consistent at low flux densities, the variable sources clearly have a higher matched fraction than the non--variables at high flux densities.

\begin{figure}
\centering
\includegraphics[scale=0.5]{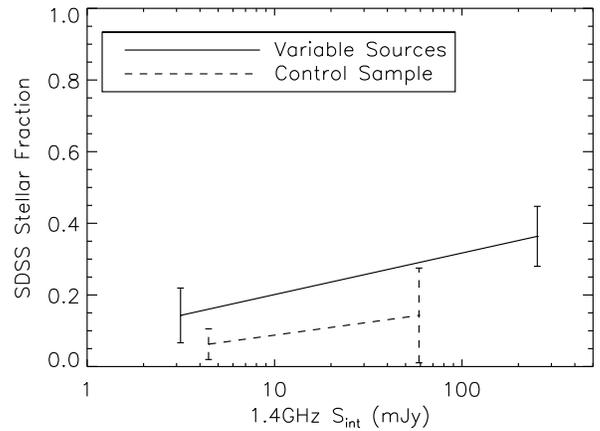}
\caption{Fraction of optically--detected radio sources that are classified as stellar based on the SDSS star/galaxy separation. Variable sources are shown by the solid line, and the non--variable control sample are shown by the dashed line.  The data points again indicate the centers of the bins.}
\label{fig:SDSSstellarfraction}
\end{figure}

The SDSS catalog includes a classification for each source based on star--galaxy separation, which differentiates between resolved and unresolved sources using the difference between the PSF magnitudes and model magnitudes (either de Vaucouleurs or exponential).  For the radio sources that were optically identified, we utilized the SDSS star--galaxy separation to learn more about the optical counterparts, taking sources classified as `stellar' to be quasar candidates.  The results are as follows:  Of the variable sources with optical counterparts, 72\% are classified as galaxies and 28\% are classified as stellar based on the star--galaxy separation, with a formal error of $\pm$6\%.  For the control sample, we find that the identified sources are composed of 92\% galaxies and 8\% stellar sources ($\pm$4\%).  These results are displayed along with the SDSS match fractions in Table \ref{tab:sdss}.  

These results demonstrate that variable sources are more likely to be quasars than non--variable sources.
They also show that, despite this fact, a significant majority of the variable sources appear to be extended in the optical. 
This could indicate that these sources are AGN in galaxies as opposed to quasars.  
We should caution that for fainter sources, the SDSS star--galaxy separation is not as reliable as for bright sources.  In addition, the distinction may not be so clear--cut for these deep optical data.  Many low--z quasars that were point sources in the single epoch data are now extended sources, because the stacking brings out the galaxy.
However, we can still learn something about the underlying source population by looking at the flux dependence with respect to the control sample.
To see if the stellar fraction depends on flux density, we plot stellar fraction as a function of flux density in Figure \ref{fig:SDSSstellarfraction}.  We split the data into two bins, and the plotted points represent the centers of those bins.  
Although the stellar fraction increases with flux density, it is below 40\% for both bins.  We therefore find that optically--extended sources make up the majority of variable sources, regardless of flux density, but a significant minority of sources are still quasars that outshine the host galaxy.
We also see that the stellar fraction of the variable sources 
is larger than that of the non--variable sources for both high and low flux densities, although the error bars are large enough that this result is not statistically significant.

	

To explore the quasar fraction further, we also matched the variable and control samples to a sample of spectroscopically--identified quasars from the SDSS.  We used the fifth edition of the SDSS quasar catalog \citep{2010AJ....139.2360S}, which is based on the Seventh Data Release (DR7).  The catalog contains 2083 spectroscopically--confirmed quasars in the area covered by the variable source sample.  We find that 9 out of 89 variable radio sources match to a spectroscopically--confirmed quasar within 1$^{\prime\prime}$, or 10\% $\pm$ 3\% (Table \ref{tab:qsos}).  For the non--variable control sample, the percentage is 3\% $\pm$ 2\%.  
Although limited by small--number statistics, we find that
the variable sources in our sample are, again, 
more likely to be quasars than the radio sources in the control sample.  
 
As a final comparison, we matched the variable and non--variable control samples to a larger catalog of photometrically--selected quasars.  The SDSS sample of photometrically--selected quasars comes from the work of \citet{Richards:2009p67}.  The catalog is expected to be $\sim$80\% efficient and contain 850,000 bona fide quasars in the area covered by SDSS DR6.  There are approximately 7,000 photometrically--selected quasar candidates in the area covered by our observations.  Matching against this catalog, we find that 13\% $\pm$ 4\% of the variable sources correlate with photometrically--selected quasars, versus 2\% $\pm$2\% of the control sample sources.  Here, the quasar fraction of the variables is (again) higher than that of the control sample  (though we caution again that we are dealing with small number statistics).   The control sample actually matches to the photometric catalog at a lower rate than it matches to the spectroscopic catalog even though the photometric catalog contains many of the spectroscopically--confirmed quasars.  This is because the spectroscopic catalog contains some quasars that are not in the photometric catalog, presumably because they exhibit abnormal colors for quasars and were targeted for spectra for some other reason.  The control sample matched to several of these quasars, while not matching to any purely photometrically--selected quasars, resulting in a lower matched fraction than when it was matched against the spectroscopic catalog.

Perhaps a more interesting observation is that, despite having 3.5 times the number of quasar candidates as the spectroscopic catalog, the photometric quasar catalog produces only $\sim$50\% more matches with the variable sample.  This is the exact same effect seen by \citet{2011AJ....142....3H} when matching the entire Stripe 82 catalog to both the spectroscopic and photometric quasar catalogs. There, it was shown that the two samples are actually entirely consistent to roughly r $=$ 20, meaning that the discrepancy lies entirely with faint sources.  The likely explanation has to do again with the SDSS targeting pipeline.  The SDSS targets sources for spectra for a number of reasons, and the selection is quite efficient.  This leaves the photometric catalog to fill in mainly those objects without radio detections, which are presumably more numerous.


\subsection{Morphology}
\label{morph}

The higher angular resolution of the Epoch III Stripe 82 data means that we can investigate in more detail the morphology of both the newly--identified and the previously--known variable sources.  In order to characterize the morphology of the sources, we follow \citet{deVries:2004p2565} and define the following six morphological classes:  (PS) -- Isolated point source; (CJ) -- ``core--jet" morphology, either as two separate components or as a single component with a clear elongation; (CL) -- ``core--lobe" morphology, where the core is surrounded by two distinct lobe components, which are not variable; (CH) -- ``core--halo" morphology, consisting of a point source core surrounded by a diffuse radio halo; (CX) -- complex source, consisting of multiple components; and lastly, (HS) -- hot--spot variability.  These morphological classes are indicated for the new and previously--identified variable sources in Tables \ref{tab:EpochIIIvariables} and \ref{tab:FZDvariables}.

Of the 58 previously--known variables in the area covered by the new Stripe 82 radio data, 45 were classified (based on the FIRST imaging) as having PS morphology, and 13 fell into other classes \citep[8 CJ, 4 CL, and 1 CX;][]{deVries:2004p2565}.
With the higher--resolution Stripe 82 data, we resolve four of the CJ sources enough to now see a CL morphology, as well as one of the CL sources into a CX morphology.  For three of the sources previously classified as PS, we resolve the radio source into CJ morphology.  Only one source moved into a simpler morphological class, and that is the source previously classified as CX, which now appears to be a PS with an unrelated neighbor.  
The breakdown we find is therefore 42 PS (70.7\%), 7 CJ (17.2\%), 7 CL (10.3\%), 1 CX (1.7\%), and no CH or HS.   One source was not detected in the Stripe 82 catalog and therefore was too faint to classify.  All of the variability is consistent with being due to the radio core/AGN component.  

We next look at the morphology of the 89 variable sources identified in this work by the addition of a third epoch of radio data.  
Note that 14 were identified previously by \citet{deVries:2004p2565}, but 75 are newly discovered.  
We find that four sources fall into the CJ class ($\sim$5\%), and five qualify as CL ($\sim$5\%).  No sources show any of the more unusual morphologies (CH, CX, or HS).  Therefore, a full 90\% of our variable sample are simple point sources.  

\begin{figure*}
\centering
\includegraphics[scale=0.85]{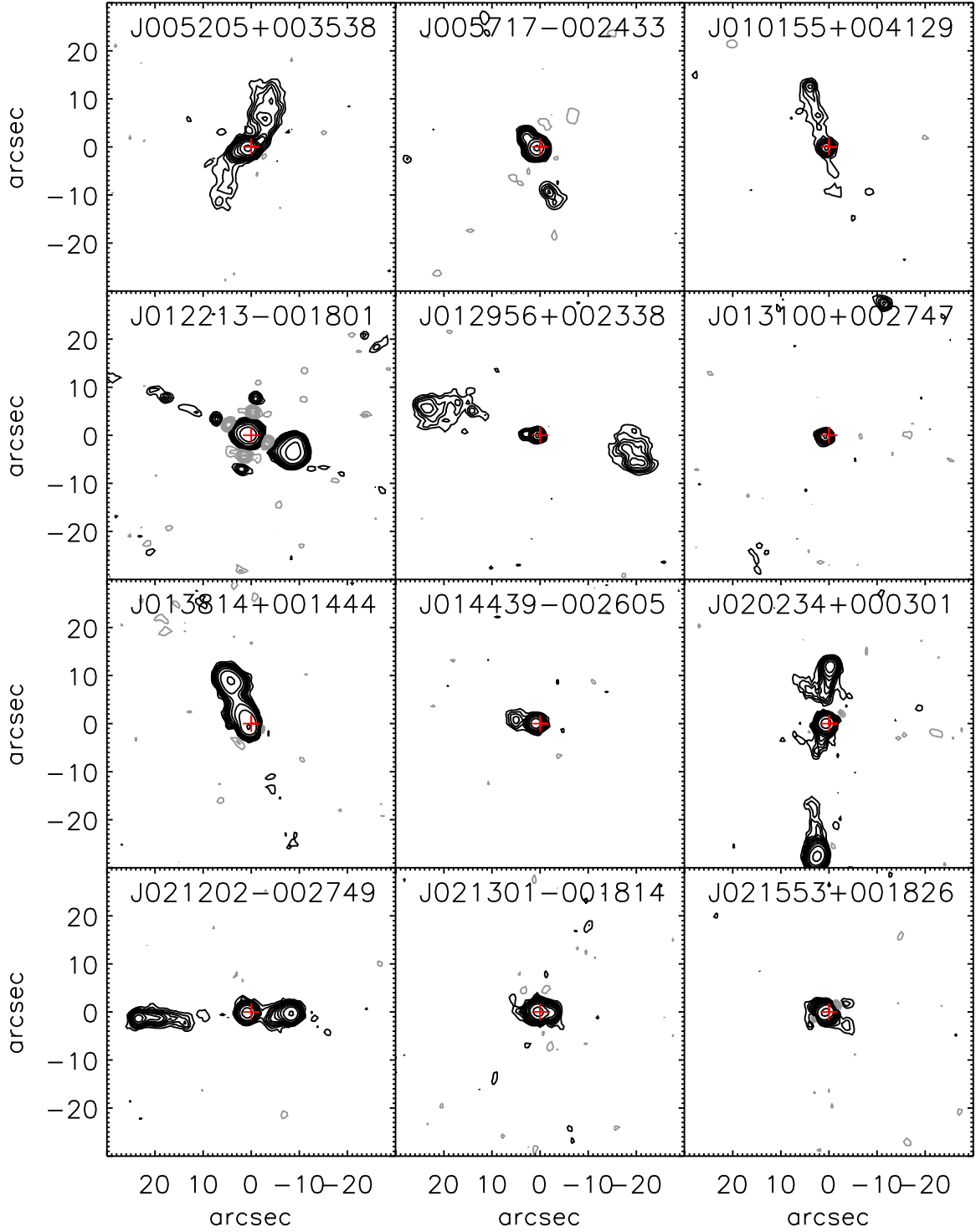}
\caption{Contour plots of the variable sources with complex morphology, using data from the Stripe 82 survey.  The cross indicates the position of the variable component .  Contours are 60 $\mu$Jy $\times$ $\pm$(3,5,7,9,12,15,20,30,50,100,300,1000), where black contours are positive and grey contours are negative.}
\label{fig:contourplots}
\end{figure*}

\begin{figure*}
\centering
\includegraphics[scale=0.85]{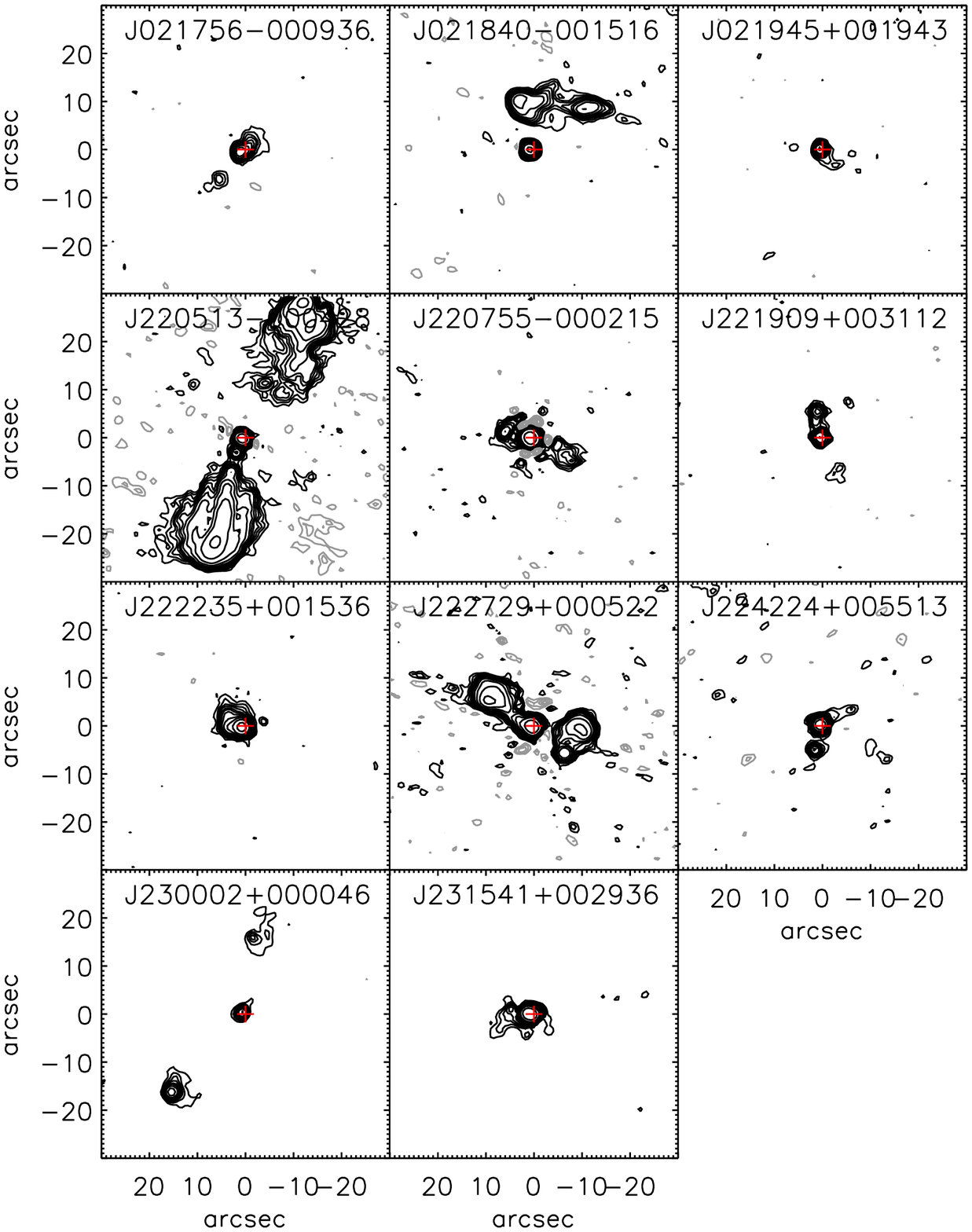}
\begin{center}
Fig. \ref{fig:contourplots} (continued)
\end{center}
\end{figure*}


Figure~\ref{fig:contourplots} shows contour plots for each of the previously--known and newly--identified variable sources that have non--PS morphology.  The images are taken from the high resolution Stripe 82 radio data.  The position of the variable component 
is indicated with a cross.  
Some of the images (e.g., J012213--001801 and J220755--00215) exhibit strong negative (and positive) contours characteristic of incompletely--cleaned sidelobes and missing short spacings.
These features are clearly image artifacts and were accounted for in the morphologoical classification.
Further notes on individual variable sources, including the complex source J222729+000522 (Figure~\ref{fig:J222729}) can be found in the Appendix. 
Any references to optical matching refer to the deep coadded SDSS catalog \citep{2011arXiv1111.6619A}.

\begin{figure}
\centering
\includegraphics[scale=0.55]{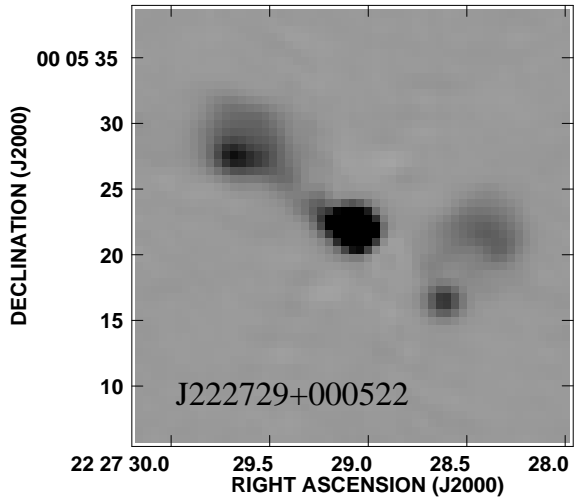}
\caption{Greyscale image (zoomed--in) of complex source J222729+000522.}
\label{fig:J222729}
\end{figure}


\section{Discussion}
\label{discussion}

Our study of long--term AGN variability extends the work of \citet{deVries:2004p2565} by introducing a new epoch of high--quality, high--resolution data.  One thing that the addition of a third epoch allows us to do is to constrain the characteristic timescale of variability for these sources, providing insight into the mechanism for variability as well as helping to identify the type of source.  For example, there is some evidence that QSO have longer characteristic timescales than BL Lac objects \citep{Hughes:1992p469}. 
Because we are measuring timescales in our reference frame instead of transforming to the source frame, we are introducing 
an additional source of 
dispersion into the data.  However, no known correlation exists between redshift and timescale beyond $1+z$, 
and intrinsic beaming is likely to be a more important effect. 

We find that extending the observing span from 7 to 14 years produces $<$50\% more variables.  
From this, we can infer an upper limit on the characteristic timescale for AGN variability of 14 years.   
We compare this result to the long--term monitoring campaign carried out by \citet{Aller:1992p16} on the University of Michigan 26m paraboloid.  \citet{Hughes:1992p469} performed a structure function analysis on the QSO and BL Lac objects in the sample, finding mean characteristic timescales of 1.95 and 2.35 years, for BL Lacs and QSOs, respectively, though with very broad distributions.  They found that the majority of sources with timescales greater than 10 years were QSOs.  In a longer--term analysis of the same data, \citet{2003ApJ...586...33A} reported that the variability of the galaxies in the sample consisted of slower variations, often only identifiable over a few decades.  They attributed this to relatively long characteristic timescales in combination with low luminosity of the AGN relative to that of the extended structure, requiring unusually bright events for detection.  Both papers were based on a very bright ($S_{\rm 5GHz}$ $>$ 1.3 Jy) sample of objects, however.   Our study is the first study to try to determine the characteristic sample of such a faint (and blind) sample.  
Additional epochs are required in order to constrain the characteristic timescale further. 

The fraction of sources that are variable at the milljansky level is another unknown. At high flux density levels, variable sources typically correspond to flat--spectrum AGN such as blazars and flat--spectrum QSOs. However, there is a general consensus that the nature of the radio source population changes with decreasing flux density. The source counts flatten below a few millijanskies \citep{1985ApJ...289..494W}, which is generally taken as evidence that a new type of source dominates below that level.  The exact composition of the population and how it depends on flux density is still hotly debated \citep[e.g.][]{2010ApJ...714.1305S, 2009ApJ...694..235P, 2009MNRAS.397..281I}. 
What is clear is that around a flux density of $\sim$1 mJy, flat--spectrum AGN make up only a small percentage of the population, with some mixture of star--forming galaxies and steep--spectrum sources making up the remainder \citep[e.g.][]{2000ExA....10..419H, 2011MNRAS.411.1547P}. 
One might therefore naively expect the variable fraction of sources to decrease with flux density as well.  Our results, on the other hand, imply that the fraction of sources that are variable stays constant down to low flux densities, and possibly even increases slightly.

\citet{2011ApJ...742...49T} have published the largest sample of millijansky variable sources to date, but they do not investigate how the variable fraction depends on flux density. The only other blind variability survey of sufficient size and depth to compare to is the work of \citet{deVries:2004p2565} on the two FIRST epochs.  However, they specifically adjusted their selection criteria to force the variable fraction to be constant, so a comparison is out of the question.  
This is therefore the first large--scale study (to our knowledge) investigating the dependence of radio variability on flux density down to the millijansky level.
We conclude that the variable fraction does not decrease at faint flux densities despite the changing composition of radio sources. 
This may imply that the variable fraction will remain constant (or even increase) down into the sub--millijansky regime for next generation surveys like that planned by ASKAP and MeerKAT \citep[although c.f.][]{2003ApJ...590..192C}.


In addition to variability, the existence of multiple epochs of data allows us to look for the more rare, transient events as well.  Because of the resolution bias, we were only able to search in the Stripe 82 epoch, but we found no evidence for any such events.  \citet{deVries:2004p2565} also reported finding no evidence for transients down to 2 mJy in their search over the larger area covered by both FIRST epochs.   

Recently, \citet{Becker:2010p157} showed that Galactic sources are much more highly variable than extragalactic sources.  We have confirmed this result, finding that only 1\% of extragalactic sources have a fractional variability $>$3, while 44\% of Galactic sources vary by this much.  The distribution of variability amplitude seen in our sample is roughly consistent with the extragalactic sample of \citet{deVries:2004p2565} used in the original comparison, although we see slightly more variability of intermediate amplitude.  
We consider our results to be more reliable than the \citet{deVries:2004p2565} results, as our images are much higher quality than the FIRST images used exclusively there.  

We saw in Section \ref{optcounterparts} that the overall optical identification rate of the variable sources is higher than that of the non--variable control sample, 
and that the effect increases at higher flux densities.
Our match rates for the higher flux density bin agree with the overal match rates \citet{deVries:2004p2565} found for their brighter (on average) variable sample -- they reported that 82\% of the variables were identified, versus only 38\% of the control sample.
Our results imply that our variable sample contains a higher fraction of quasars than our non--variable sample, 
particularly at higher flux densities.
As quasars are unresolved, an imaging survey will detect them more easily than extended sources of the same (integrated) magnitude.
As the unidentified radio sources are most likely obscured AGN, our results also imply that variable sources are less likely to be obscured AGN than non--variable sources, though the effect decreases at lower flux densities.  

Of the optically--identifiable variable sources, we find that a majority match to extended sources (galaxies), as opposed to point sources (quasars).  
In contrast, a full half of the variables studied by \citet{deVries:2004p2565} were optical point sources.
Our sample is fainter than the de Vries et al. sample in the radio, on average, which could explain some of the discrepany.  
Not only are fainter sources more likely to be galaxies, but the SDSS star--galaxy separation is less reliable at low flux densities.  
However, when we calculate the stellar fraction as a function of flux density, it never approaches the higher percentages seen in \citet{deVries:2004p2565} even for high values of flux density.  
This may be largely because the coadded optical data we compare to go much deeper, 
allowing us to identify more of the counterparts (not just the bright quasars),
but also
allowing the host galaxies of many low--z quasars to be detected, and blurring the line between traditionally--defined ``quasars" and generic AGN in galaxies.  

We also compare these results to the optical point source fraction of a control sample, defined to be non--variable in the radio.
We find that the variable sources are more likely to match to point sources (quasars) than control sample sources.  
This is further confirmed by matching both samples to actual spectroscopic and photometric quasar catalogs (though with the added caveat that the results suffer from small number statistics).
This result is consistent with the results of \citet{deVries:2004p2565}, who also used the SDSS star--galaxy separation to classify their identifiable fraction
and reported a significantly higher stellar fraction in the variables than in the control sample.




Upon examining the morphology of the previously--identified variable sources using the new, higher--resolution data, we find that 26\% have structure that is more complex than a simple point source.  
We identify new jet/lobe components in a number of the variables, resulting in an increased fraction of complex sources over the 21.8\% reported by \citet{deVries:2004p2565}.  
This confirms their finding of a strong correlation between radio variability and jet activity for this sample of radio variables.  
Furthermore, as we resolve four of the sources previously classified as CJ enough to see radio lobes, we believe that deeper data of the same angular resolution would find than an even higher fraction of variables actually reside in the cores of FR II radio galaxies.   
	
We also examine the morphology of our sample of 89 variable sources.  According to the unified model of AGN, extended emission would arise from the radio lobes, which are not expected to be variable.  All of the variability we see is consistent with originating from the core, which is consistent with the model.  A full 90\% of the variables are, in fact, simple point sources, as opposed to being cores of FRIIs, or showing jet--like features.  This is significantly different from the results for the previously--identified variables, and may be due to a couple of reasons.  
For one thing, our point source criterion results in a stricter cut due to our higher resolution, and we are therefore more biased against sources which are extended due to the presence of jets. 
The higher angular resolution data may also resolve out some of the extended structure.  
Finally, our sample is fainter, on average, than the de Vries et al.\ sample.  Although the Epoch III data are three times more sensitive, the factor of three increase in angular resolution means that diffuse components are spread out over nine times the area.  This would make diffuse components of the same fractional strength much harder to detect.  





\section{Conclusions}
\label{conclusions}

We carried out a blind survey for extragalactic radio variability at the millijansky level by comparing two epochs of data from the FIRST survey with a new 1.4 GHz survey of SDSS Stripe 82.  
The three epochs, which we refer to as I, II and III, are spaced seven years apart and cover an overlapping area of 60 deg$^{2}$.  Our main results are the following:

\begin{itemize}


\item We identified 89 variable sources with a variability amplitude above 30\%.   Of these, 14 sources were previously reported by \citet{deVries:2004p2565} and 75 are newly--identified.  Due to the resolution bias, our sample is restricted to those sources that increased in flux density in Epoch III.  Assuming that an equal number of sources increase and decrease in flux density, we would expect approximately 178 variable sources in total, or $\sim$3 variables deg$^{-2}$.   

\item We found no evidence for transient phenomena in the new Stripe 82 data.

\item The fraction of radio sources that are variable appears to remain constant down to the millijansky level, possibly even increasing slightly at low flux densities.
This is despite the fact that the fraction of flat--spectrum (i.e. canonically variable) sources decreases with decreasing flux density.  
This may imply that next generation radio surveys with telescopes like ASKAP and MeerKAT will see a constant or even increasing variable fraction down into the submillijansky regime.

\item By comparing the distribution of fractional variability against that of a Galactic sample, we confirmed that Galactic sources tend to be more highly variable than extragalactic sources.  While 44\% of Galactic sources vary by more than 200\%, only 1\% of extragalactic sources have variability amplitudes this large.

\item When we compared the number of variables discovered over different lengths of time, we found that increasing the observing interval from 7 to 14 years only resulted in the detection of $<$50\% more sources.  From this, we inferred an upper limit to the mean characteristic time of AGN variability of 14 years.  

\item We found that the overall rate of SDSS optical identification is higher for the variable sources than for a 
flux--matched non--variable control sample, 
though the effect depends on flux density.
This implies that the variable sample contains a higher fraction of quasars than the non--variable sample, a result which is
confirmed by both the SDSS star--galaxy classification and comparisons to spectroscopic and photometric quasar catalogs.
The majority of variable sources at all flux densities appear to be extended galaxies in the optical, implying either that 
quasars are not the dominant contributor to the variability of this sample, or that
the deep optical data allow us to detect the host galaxies of many low--z quasars,
blurring the line between traditionally--defined ``quasars" and generic AGN.


\item We used the new epoch of high--resolution data to characterize the radio morphology of the variable sample, concluding that over 90\% of our variable sample are stand--alone point sources.  This sample does not appear to exhibit the strong correlation with radio jets seen in the only other blind survey of millijansky AGN variability.  The resolution bias is likely partly to blame.



\end{itemize}

While this work serves as a starting point for the study of faint AGN variability, clearly more data are needed if we wish to better understand the variable population and characterize the variability.  Deeper, high resolution data are needed to identify the population, and additional epochs are essential to constrain the characteristic timescale and mechanism for variability.  
Nevertheless, this survey gives us a small taste of what is to come with variability campaigns on upcoming telescopes like ASKAP and MeerKAT.


\acknowledgements
JAH acknowledges the support of NRAO Grant GSSP08--0034, a UC Davis Graduate Block Grant Fellowship, and Grant HST--GO--10412.03--A from the Space Telescope.  
RHB acknowledge the support of the National Science Foundation under grant AST 00--98355.  
The work by RHB was partly performed under the auspices of the U.S. Department of Energy by Lawrence Livermore National Laboratory under Contract DE--AC52--07NA27344.  
RLW acknowledges the support of the Space Telescope Science Institute, which is operated by the Association of Universities for Research in Astronomy under NASA contract NAS5--26555.  
GTR acknowledges the support of NSF grant AST-1108798.

The National Radio Astronomy Observatory is a facility of the National Science Foundation under cooperative agreement by Associated Universities, Inc.


\appendix
\section{Notes on individual variable sources}

\textbf{J005205+003538} (CL):  Previously known to be variable.  Diffuse lobes are now visible on either side of the core.  Although neither component appears in the Stripe 82 catalog, we believe this fits better in the CL category than the CJ category, as two diffuse lobes are clearly visible.  The central (variable) component corresponds to an optical QSO at $z = 0.399$ with a magnitude of r $=$ 16.3.  

\textbf{J005717-002433}  (CJ):  Previously known to be variable.  The contour plot shows that the FIRST source has been resolved into three components, although only the central source shows up in the catalog.  There is a core, a bright point source connected to the NE, and a separate extended component to the SW.  Because the relationship of the components is unclear, we classify this as a core--jet system.  
The central (variable) source matches to a QSO at $z = 2.712$. 

\textbf{J010155+004129}  (CJ):  A newly--identified variable source.  The higher--resolution Stripe 82 data show a diffuse, jet--like stream connecting the central source to a hotspot in the north.  This stream consists of three components in the Stripe 82 catalog: a component 7$^{\prime\prime}$ away with a peak flux density of 0.4 mJy beam$^{-1}$ and an integrated flux density of 2.3 mJy; an elongated component 10$^{\prime\prime}$ away with a peak flux density of 0.4 mJy beam$^{-1}$ and an integrated flux density of 6.3 mJy; and a third component 12$^{\prime\prime}$ away with a peak flux density of 0.6 mJy beam$^{-1}$ and an integrated flux density of 2.6 mJy.  The central (variable) radio component corresponds to an optical QSO at $z = 0.649$.   


\textbf{J012213-001801}  (CL):  Previously known to be variable.  The FIRST image showed a core and two lobes, one (the northeastern lobe) much fainter than the other (peak flux density of 1.7 mJy beam$^{-1}$). In the Stripe 82 data, the northeastern lobe is (mostly) resolved out and fainter than the nearby sidelobes. It does not make it into the Stripe 82 catalog.
The central (variable) component has a peak flux density of 331.3 mJy beam$^{-1}$ and an integrated flux density of 405.0 mJy.
It coincides with a photometrically--selected $z = 1.325$ UV--excess (UVX) 
quasar candidate \citep{Richards:2004p257}.  
The southwest component has a peak flux density of 38.2 mJy beam$^{-1}$ and an integrated flux density of 116.2 mJy and does not have an optical counterpart.  

\textbf{J012956+002338}  (CL):  A newly--identified variable source.  This radio--double has an FRI morphology, with the lobes terminating in hot spots and a faint jet--like feature visible near the north--eastern lobe.  
The north--eastern lobe is visible in the image, but has been resolved out somewhat and does not appear in the catalog.
The south--western lobe has been resolved into three distinct components: a component 19$^{\prime\prime}$ away with a peak flux density of 1.0 mJy beam$^{-1}$ and an integrated flux density of 1.9 mJy; a component 22$^{\prime\prime}$ away with a peak flux density of 0.8 mJy beam$^{-1}$ and an integrated flux density of 1.1 mJy; and a third component 23$^{\prime\prime}$ away with a peak flux density of 1.2 mJy beam$^{-1}$ and an integrated flux density of 2.1 mJy.  The variable component (core) corresponds to a QSO at $z = 1.079$.

\textbf{J013100+002747}  (CL):  A newly--identified variable source.  
The FIRST data show only the central point--source, but the Stripe 82 data show evidence of a double--lobed structure, although the southeastern component is only detected at 1$\sigma$ significance.  
The central (variable) component matches (at 0.07$^{\prime\prime}$) to a galaxy of magnitude r $=$ 21.7 in the deep optical catalog.  
The northwestern component has a peak flux density of 0.8 mJy beam$^{-1}$ and an integrated flux density of 1.4 mJy.  
This component matches (within 0.04$^{\prime\prime}$) to a stellar source of r $=$ 24.2 in the deep optical catalog, suggesting that it may be unrelated.   
In this case, the morphological classification would be PS.

\textbf{J013814+001444}  (CJ):  Previously known to be variable.  The FIRST source has been resolved into two components.  The central (variable) component has a peak flux density of 18.9 mJy beam$^{-1}$ and an integrated flux density of 62.1 mJy.  The northeastern component has a peak flux density of 6.5 mJy beam$^{-1}$ and an integrated flux density of 33.5 mJy.  Neither component has an optical counterpart in the deep optical data.  

\textbf{J014439-002605}  (CJ):  A newly--identified variable source.  The Stripe 82 data show two components which were previously unresolved in the FIRST survey.  The central (variable) component has a peak flux density of 14.7 mJy beam$^{-1}$ and an integrated flux density of 15.4 mJy.  The component 4$^{\prime\prime}$ to the east has a peak flux density of 0.7 mJy beam$^{-1}$ and an integrated flux density of 2.0 mJy.  The central source matches (within 0.03$^{\prime\prime}$) to a stellar source of r $=$ 21.1 in the deep optical data.  The component to the east has no optical match.  

\textbf{J020234+000301}  (CL):  Previously known to be variable, and also found to be variable in the Stripe 82 radio data.  The three sources all have jet--like structures in the new Stripe 82 data.  The central (variable) component has a peak flux density of 41.7 mJy beam$^{-1}$ and an integrated flux density of 44.6 mJy.  The lobe 12$^{\prime\prime}$ to the north has a peak flux density of 4.6 mJy beam$^{-1}$ and an integrated flux density of 6.6 mJy.  The lobe 28$^{\prime\prime}$ to the south has a peak flux density of 29.5 mJy beam$^{-1}$ and an integrated flux density of 49.0 mJy.  The central component coincides with a $z = 0.366$ quasar.  The southern component is 2.95$^{\prime\prime}$ from an r $=$ 22.9 galaxy in the deep optical data, but since the morphology is very suggestive of a double--lobed radio galaxy, this is likely a coincidence.  

\textbf{J021202-002749}  (CL):  Previously known to be variable.  Three distinct components were already detected in the FIRST data, but the eastern lobe has now been resolved into two components of peak flux density 1.3/4.2 mJy beam$^{-1}$ and integrated flux density 6.2/4.5 mJy closer/further from the center, respectively.  The corresponding FIRST lobe had a peak flux density of 6.0 mJy beam$^{-1}$ and an integrated flux density of 9.2 mJy.  The western lobe has a Stripe 82 peak flux density of 21.4 mJy beam$^{-1}$ and an integrated flux density of 30.2 mJy, while the FIRST lobe had a peak flux density of 30.2 mJy beam$^{-1}$ and an integrated flux density of 31.8 mJy.  The deep optical imaging shows a source 0.14$^{\prime\prime}$ from the central (variable) component.  It is classified as a galaxy of magnitude r $=$ 23.5.  The other components have no optical matches.  

\textbf{J021301-001814}  (CJ):  Previously known to be variable.  Called a PS by \citet{deVries:2004p2565}, the new data begin to resolve this source, showing evidence of CJ morphology.  An optical source 0.13$^{\prime\prime}$ from the variable radio source is classified as a galaxy of r $=$ 22.2.  

\textbf{J021553+001826}  (CJ):  Previously known to be variable.  The source showed PS morphology in the FIRST survey, but the Stripe 82 data show evidence of a 2$\sigma$ extension to the south--west.  The closest optical match (at 0.1$^{\prime\prime}$) is a galaxy of magnitude r $=$ 19.6 in the deep optical catalog.  

\textbf{J021756-000936}  (CJ):  Previously known to be variable.  This source was previously one component and classified as CJ, but it has been resolved into two sources in the Stripe 82 data.   The central (variable) component has a peak flux density of 2.4 mJy beam$^{-1}$ and an integrated flux density of 6.7 mJy.  The southeast component has a peak flux density of 0.5 mJy beam$^{-1}$ and an integrated flux density of 1.4 mJy.  A faint jet--like structure extends up to the northwest of the core.  There appears to be an object approximately 1$^{\prime\prime}$ from the core component in the deep optical catalog; however, an artifact in the image resulted in its not appearing in the catalog.


\textbf{J021840-001516}  (PS):  Previously known to be variable.  This source is included in the list because \citet{deVries:2004p2565} classified it as a complex (CX) source, with three different components in an unusual configuration.  They were unclear if the various components related at all.  This multi--component source has been resolved into four separate components with the Stripe 82 data, and in the new data, the sources to the North look unrelated to the variable source.  The component 14.8$^{\prime\prime}$ to the northwest coincides (within 0.61$^{\prime\prime}$) with a r $=$ 23.7 galaxy in the deep optical data.  There appears to be a double--lobed radio galaxy associated with this source.  The variable radio source, which corresponds to a $z = 1.171$ quasar, is therefore a simple PS. 

\textbf{J021945+001943}  (CJ):  A newly--identified variable source.  The source consists of a central component with a jet--like feature extending to the south--west.  The radio source matches to an optical quasar at $z = 1.266$.   

\textbf{J220513-000428}  (CL):  A newly--identified variable source, and a clear double--lobed radio source. Three components (the core and two lobes) are already visible in the lower--resolution FIRST data. The Stripe 82 data resolves the source even further, with the southern lobe listed as two extended components with integrated flux densities of 118.2 and 49.4 mJy. The northern lobe is extended as well, with an integrated flux density of 43.7 mJy, and the (variable) core is point--like with a peak flux density of 11.5 mJy beam$^{-1}$. Within 0.1$^{\prime\prime}$ of the core is a r $=$ 21.8 object classified as a galaxy in the deep optical catalog.

\textbf{J220755-000215}  (CL):  Previously known to be variable.  The FIRST catalog shows the central source and an extremely faint point source 21$^{\prime\prime}$ to the southwest with a peak flux density of only 0.8 mJy beam$^{-1}$.  This source is likely to be a side--lobe, and does not appear in the Stripe 82 image or catalog.
In the Stripe 82 data, the central FIRST source has been resolved into two components (a central source and a lobe to the northeast with a peak flux density of 1.8 mJy beam$^{-1}$ and an integrated flux density of 3.8 mJy), and the image implies the existence of a third component to the southwest that did not make it into the catalog.  
The source therefore appears to be a double--lobed radio source.  The central (variable) component has an optical match 0.14$^{\prime\prime}$ away in the deep optical catalog that is classified as stellar with a magnitude r $=$ 21.1.

\textbf{J221909+003112}  (CL):  Previously known to be variable.  Whereas FIRST saw only one component, we see that it has now been resolved into five different components in the Stripe 82 data:
the component to the south of the central component has a peak flux density of 0.4 mJy beam$^{-1}$ and an integrated flux density of 0.5 mJy; 
a component 2.4$^{\prime\prime}$ to the north has a peak flux density of 0.7 mJy beam$^{-1}$ and an integrated flux density of 0.8 mJy; 
a component 4.7$^{\prime\prime}$ to the north has a peak flux density of 1.4 mJy beam$^{-1}$ and an integrated flux density of 2.5 mJy; 
and the point source component to the northwest has a peak flux density of 0.4 mJy beam$^{-1}$.   
It is unclear whether this last component is related.  
The central (variable) component matches (within 0.11$^{\prime\prime}$) to a r $=$ 22.4 object classified as stellar in the deep optical catalog.  
None of the other components have close matches in the deep optical catalog.  

\textbf{J222235+001536}  (CJ):  Previously known to be variable.  The source was classified by \citet{deVries:2004p2565} as having a CJ morphology, but we would have called this source a PS based only on the FIRST data.  The Stripe 82 data show more structure, giving more credence to a CJ classification.  An optical match at 0.13$^{\prime\prime}$ from the source is classified as a QSO at $z = 1.362$.  

\textbf{J222729+000522}  (CX):  Previously known to be variable.  The FIRST data show three components -- a central source and two lobes -- but the southwestern lobe has now been resolved into two separate components.    
In the FIRST data, this lobe had a peak flux density of 17.6 mJy beam$^{-1}$ and an integrated flux density of 34.5 mJy.  
In the Stripe 82 data, the southernmost component has a peak flux density of 6.1 mJy beam$^{-1}$ and an integrated flux density of 9.1 mJy.  
The northern component of that lobe has a peak flux density of 3.5 beam$^{-1}$ mJy and an integrated flux density of 24.2 mJy.  
The north--eastern lobe has a Stripe 82 peak flux density of 6.1 mJy beam$^{-1}$ and an integrated flux density of 46.3 mJy.  
The FIRST data gave a peak flux density of 28.8 mJy beam$^{-1}$ and an integrated flux density of 45.5 mJy for that lobe.  
The deep optical data shows a source 0.22$^{\prime\prime}$ from the central (variable) component.  It is classified as a quasar with $z = 1.518$.  
None of the other components show close matches. 
A greyscale image of this source is shown in Figure~\ref{fig:J222729} to give another view of the complex morphology.

\textbf{J224224+005513}  (CL):  Previously known to be variable.  The FIRST source has been resolved into two components in the high--resolution Stripe 82 data, with a third component (to the northwest) just below the detection threshold. 
The component to the south has a peak flux density of 1.2 mJy beam$^{-1}$ and an integrated flux density of 2.3 mJy.   
The deep optical data show a source 0.17$^{\prime\prime}$ from the central (variable) component that is classified as stellar with a magnitude r $=$ 21.0.  


\textbf{J230002+000046}  (CL):  A newly--identified variable source.  
The variable component corresponds to the core of a double--lobed radio galaxy.  
Neither of the lobes satisfy the variability criteria.  
The northwestern lobe had a FIRST peak flux density of 1.2 mJy beam$^{-1}$ and an integrated flux density of 2.5 mJy, and in the Stripe 82 catalog it has a peak flux density of 0.4 mJy beam$^{-1}$ and an integrated flux density of 3.0 mJy.  
The southeastern lobe had a FIRST peak flux density of 3.8 mJy beam$^{-1}$ and an integrated flux density of 4.4 mJy, and in the Stripe 82 catalog it has a peak flux density of 2.2 mJy beam$^{-1}$ and an integrated flux density of 5.1 mJy.  
None of the components have optical matches within 3$^{\prime\prime}$.  


\textbf{J231541+002936} (CJ):  Previously known to be variable, and also found to be variable in the Stripe 82 data.  
This source had a point source morphology in the FIRST data, but has now been resolved into three components.  
The central component has a peak flux density of 17.0 mJy beam$^{-1}$ and an integrated flux density of 23.8 mJy.  
An extended component 2.2$^{\prime\prime}$ to the east has a peak flux density of 0.5 mJy beam$^{-1}$ and an integrated flux density of 8.0 mJy, and a component 7.0$^{\prime\prime}$ to the southeast has a peak flux density of 0.4 mJy beam$^{-1}$ and an integrated flux density of 1.7 mJy.  
The two point source features to the south and northeast are too faint to make it into the catalog, which has a flux threshold which depends on radial distance from bright sources.  
The central (variable) component matches to an optical quasar at $z = 1.358$.


\bibliography{dissertation}
\bibliographystyle{apj}



\clearpage
\LongTables	

\tabletypesize{\scriptsize}
\setlength{\tabcolsep}{0.02in}
\begin{longtable}{llrrrrrrrrrrrcc}
\caption{Epoch III Variable Sources} \label{tab:EpochIIIvariables} \\

\hline\hline \\[0.5ex]
       & &  \multicolumn{3}{c}{\textbf{Epoch I}} & \multicolumn{3}{c}{\textbf{Epoch II}} & \multicolumn{3}{c}{\textbf{Epoch III}} &  &  &  & \\
      \multicolumn{1}{c}{\textbf{RA}} & \multicolumn{1}{c}{\textbf{Dec}} & \multicolumn{1}{c}{\boldmath{$S_{\rm pk}$}} & \multicolumn{1}{c}{\boldmath{$S_{\rm int}$}} & \multicolumn{1}{c}{\textbf{rms}} & \multicolumn{1}{c}{\boldmath{$S_{\rm pk}$}} & \multicolumn{1}{c}{\boldmath{$S_{\rm int}$}} & \multicolumn{1}{c}{\textbf{rms}} & \multicolumn{1}{c}{\boldmath{$S_{\rm pk}$}} & \multicolumn{1}{c}{\boldmath{$S_{\rm int}$}} & \multicolumn{1}{c}{\textbf{rms}} & \multicolumn{1}{c}{\textbf{Maj}} & \multicolumn{1}{c}{\boldmath{$f_{\rm var}$}} & \multicolumn{2}{c}{\textbf{Morphology}} \\
       & & \multicolumn{1}{c}{\textbf{(mJy)}} & \multicolumn{1}{c}{\textbf{(mJy)}} & \multicolumn{1}{c}{\textbf{($\mu$Jy)}} & \multicolumn{1}{c}{\textbf{(mJy)}} & \multicolumn{1}{c}{\textbf{(mJy)}} & \multicolumn{1}{c}{\textbf{($\mu$Jy)}} & \multicolumn{1}{c}{\textbf{(mJy)}} & \multicolumn{1}{c}{\textbf{(mJy)}} & \multicolumn{1}{c}{\textbf{($\mu$Jy)}} & \multicolumn{1}{c}{\boldmath{($^{\prime\prime}$)}} & & \multicolumn{1}{c}{\textbf{(FIRST)}} & \multicolumn{1}{c}{\textbf{(S82)}}\\      [0.5ex] \hline
       \\[0.5ex]
\endfirsthead
  
\multicolumn{15}{c}{{\tablename} \thetable{} -- Continued} \\[0.5ex]
\hline\hline \\[0.5ex]
       & &  \multicolumn{3}{c}{\textbf{Epoch I}} & \multicolumn{3}{c}{\textbf{Epoch II}} & \multicolumn{3}{c}{\textbf{Epoch III}} &  &  &  & \\  [0.5ex] 
      \multicolumn{1}{c}{\textbf{RA}} & \multicolumn{1}{c}{\textbf{Dec}} & \multicolumn{1}{c}{\boldmath{$S_{\rm pk}$}} & \multicolumn{1}{c}{\boldmath{$S_{\rm int}$}} & \multicolumn{1}{c}{\textbf{rms}} & \multicolumn{1}{c}{\boldmath{$S_{\rm pk}$}} & \multicolumn{1}{c}{\boldmath{$S_{\rm int}$}} & \multicolumn{1}{c}{\textbf{rms}} & \multicolumn{1}{c}{\boldmath{$S_{\rm pk}$}} & \multicolumn{1}{c}{\boldmath{$S_{\rm int}$}} & \multicolumn{1}{c}{\textbf{rms}} & \multicolumn{1}{c}{\textbf{Maj}} & \multicolumn{1}{c}{\boldmath{$f_{\rm var}$}} & \multicolumn{2}{c}{\textbf{Morphology}} \\
       & & \multicolumn{1}{c}{\textbf{(mJy)}} & \multicolumn{1}{c}{\textbf{(mJy)}} & \multicolumn{1}{c}{\textbf{($\mu$Jy)}} & \multicolumn{1}{c}{\textbf{(mJy)}} & \multicolumn{1}{c}{\textbf{(mJy)}} & \multicolumn{1}{c}{\textbf{($\mu$Jy)}} & \multicolumn{1}{c}{\textbf{(mJy)}} & \multicolumn{1}{c}{\textbf{(mJy)}} & \multicolumn{1}{c}{\textbf{($\mu$Jy)}} & \multicolumn{1}{c}{\boldmath{($^{\prime\prime}$)}} & & \multicolumn{1}{c}{\textbf{(FIRST)}} & \multicolumn{1}{c}{\textbf{(S82)}}\\      [0.5ex] \hline       
       \\[0.5ex]
\endhead  
  
\\[0.2ex] \hline \\[0.5ex]
\multicolumn{15}{l}{{(Continued on the next page\ldots)}} \\
\endfoot

\\[0.2ex]  \hline \hline
\endlastfoot

   \phm{*}00 41 52.484  & +00 03 02.16  &    4.01  &   3.92  &  134  &   3.34  &   3.28  &  135  &   3.88  &   4.34  &   48  &  0.68  &  1.30  & PS  & PS  \\
 \phm{*}00 42 02.303  & $-$00 16 06.35  &    5.43  &   5.20  &  146  &   5.92  &   5.89  &  149  &   6.04  &   8.12  &   57  &  1.33  &  1.50  & PS  & PS  \\
 \phm{*}00 43 37.907  & $-$00 12 44.20  &    1.27  &   1.12  &  136  &   2.24  &   2.69  &  136  &   3.82  &   3.95  &   94  &  0.43  &  3.11  & PS  & PS  \\
 \phm{*}00 46 22.696  & $-$00 10 22.33  &    7.80  &   7.95  &  131  &   8.49  &   8.55  &  130  &  10.04  &  10.28  &  168  &  0.30  &  1.32  & PS  & PS  \\
   \phm{*}00 47 08.920  & +00 06 03.02  &    4.03  &   3.98  &  152  &   4.34  &   3.48  &  154  &   4.08  &   5.43  &   79  &  1.52  &  1.35  & PS  & PS  \\  
   \phm{*}00 52 06.443  & +00 14 10.16  &    1.93  &   2.37  &  139  &   2.00  &   1.13  &  142  &   2.43  &   2.81  &   50  &  0.83  &  1.46  & PS  & PS  \\
   \phm{*}00 57 22.306  & +00 16 44.67  &    3.01  &   3.18  &  241  &   3.53  &   2.99  &  234  &   3.18  &   4.40  &  168  &  1.19  &  1.46  & PS  & PS  \\
 \phm{*}01 00 29.413  & $-$00 24 24.05  &    1.35  &   1.72  &  139  &   2.25  &   1.94  &  144  &   2.00  &   2.52  &   54  &  0.95  &  1.87  & PS  & PS  \\
   \phm{*}01 00 33.502  & +00 22 00.17  &   18.67  &  18.90  &  165  &  18.64  &  18.92  &  170  &  19.84  &  24.45  &   96  &  1.13  &  1.31  & PS  & PS  \\  
 \phm{*}01 01 48.758  & $-$00 06 46.92  &    2.63  &   3.16  &  142  &   2.11  &   4.07  &  141  &   2.51  &   3.10  &   58  &  1.04  &  1.47  & PS  & PS  \\
 \phm{*}01 01 51.209  & $-$00 02 32.85  &    3.17  &   2.31  &  151  &   3.24  &   2.41  &  150  &   3.17  &   4.46  &   62  &  1.32  &  1.41  & PS  & PS  \\
   \phm{*}01 01 55.997  & +00 41 29.20  &    4.27  &   5.13  &  141  &   3.33  &   3.40  &  148  &   3.59  &   4.56  &   58  &  1.02  &  1.37  & CJ  & CJ  \\
   \phm{*}01 05 59.953  & +00 19 04.29  &    3.42  &   3.13  &  140  &   4.52  &   4.10  &  141  &   4.45  &   5.32  &   51  &  0.86  &  1.56  & PS  & PS  \\
   \phm{*}01 08 28.579  & +00 03 12.57  &    4.01  &   4.43  &  143  &   4.45  &   4.93  &  145  &   4.61  &   5.52  &   58  &  1.00  &  1.38  & PS  & PS  \\  
  \**01 11 06.789  & +00 08 46.51  &    3.43  &   3.67  &  147  &   5.40  &   4.84  &  149  &   3.80  &   5.09  &   67  &  1.37  &  1.57  & PS  & PS  \\  
 \phm{*}01 12 48.603  & $-$00 17 24.72  &   17.79  &  19.22  &  148  &  16.86  &  17.48  &  149  &  20.93  &  28.83  &   64  &  1.45  &  1.71  & PS  & PS  \\  
 \phm{*}01 14 34.309  & $-$00 25 59.70  &    3.02  &   3.18  &  143  &   3.07  &   2.48  &  143  &   3.40  &   3.94  &   56  &  0.75  &  1.30  & PS  & PS  \\
 \phm{*}01 21 57.479  & $-$00 12 27.02  &    2.35  &   2.20  &  243  &   3.27  &   2.64  &  170  &   3.86  &   4.37  &  209  &  0.89  &  1.86  & PS  & PS  \\
 \phm{*}01 22 47.743  & $-$00 02 56.83  &    1.98  &   1.36  &  230  &   3.66  &   3.57  &  177  &   2.78  &   3.45  &  128  &  1.08  &  1.85  & PS  & PS  \\
   \phm{*}01 29 56.720  & +00 23 38.38  &    2.82  &   3.37  &  140  &   3.14  &   2.88  &  139  &   3.92  &   4.09  &  176  &  0.81  &  1.45  & CL  & CL  \\
   \phm{*}01 31 00.177  & +00 27 47.11  &    1.53  &   0.88  &  158  &   1.36  &   0.58  &  156  &   2.06  &   2.84  &  123  &  1.37  &  2.10  & PS  & CL  \\
   \phm{*}01 31 51.644  & +00 07 41.29  &    1.47  &   1.34  &  142  &   1.67  &   1.58  &  140  &   3.25  &   3.55  &   68  &  0.73  &  2.41  & PS  & PS  \\
 \phm{*}01 32 20.294  & $-$00 00 54.17  &    1.11  &   1.06  &  143  &   1.79  &   1.46  &  144  &   1.55  &   2.02  &   80  &  1.35  &  1.82  & PS  & PS  \\
   \phm{*}01 33 16.865  & +00 02 17.15  &    1.22  &   0.64  &  138  &   1.46  &   1.40  &  134  &   1.84  &   2.16  &  130  &  0.83  &  1.77  & PS  & PS  \\
   \phm{*}01 34 00.142  & +00 09 31.37  &    1.71  &   1.10  &  204  &   1.59  &   1.55  &  141  &   2.61  &   2.75  &  439  &  0.79  &  1.73  & PS  & PS  \\
 \phm{*}01 44 39.861  & $-$00 26 05.31  &   11.14  &  13.66  &  135  &  12.68  &  14.69  &  138  &  14.68  &  15.38  &   56  &  0.47  &  1.38  & PS  & CJ  \\
  \phm{*}01 52 14.399  & +00 15 02.01  &    1.35  &   1.04  &  137  &   1.20  &   1.14  &  138  &   1.89  &   2.30  &  231  &  0.98  &  1.91  &  PS  & PS  \\
   \phm{*}01 59 40.965  & +00 21 37.18  &    3.33  &   2.82  &  140  &   3.70  &   3.30  &  141  &   4.82  &   5.04  &   55  &  0.50  &  1.51  & PS  & PS  \\
 \phm{*}02 00 06.382  & $-$00 10 48.18  &    1.24  &   0.61  &  149  &   1.85  &   1.16  &  149  &   1.99  &   2.78  &   62  &  1.44  &  2.24  & PS  & PS  \\
   \phm{*}02 01 26.777  & +00 01 45.54  &    5.81  &   6.31  &  149  &   4.65  &   4.05  &  149  &   5.83  &   7.20  &   63  &  1.23  &  1.55  & PS  & PS  \\
   \phm{*}02 01 55.957  & +00 32 13.90  &    1.23  &   1.67  &  144  &   1.78  &   1.31  &  146  &   1.62  &   2.19  &   58  &  1.45  &  1.78  & PS  & PS  \\
   \**02 02 34.322  & +00 03 01.83  &   39.41  &  44.01  &  141  &  30.99  &  35.42  &  142  &  41.74  &  44.64  &  436  &  0.59  &  1.44  & CL  & CL  \\  
 \phm{*}02 03 15.893  & $-$00 14 32.54  &    1.83  &   1.82  &  150  &   2.54  &   1.62  &  152  &   2.52  &   3.30  &   75  &  1.28  &  1.80  & PS  & PS  \\
 \phm{*}02 14 39.295  & $-$00 24 05.36  &    5.25  &   5.15  &  137  &   3.96  &   3.66  &  140  &   4.69  &   5.23  &   54  &  0.68  &  1.33  & PS  & PS  \\
 \phm{*}02 14 54.909  & $-$00 21 00.89  &    3.08  &   2.98  &  136  &   2.58  &   2.19  &  136  &   3.49  &   3.65  &   52  &  0.39  &  1.41  & PS  & PS  \\
   \phm{*}02 19 45.435  & +00 19 43.31  &    2.83  &   3.22  &  141  &   2.84  &   3.12  &  143  &   3.32  &   3.84  &   80  &  0.95  &  1.36  & PS  & CJ  \\
 \phm{*}22 04 55.826  & $-$00 01 47.25  &    1.93  &   1.15  &  153  &   2.83  &   2.38  &  152  &   2.45  &   3.28  &  258  &  1.97  &  1.70  & PS  & PS  \\
\phm{*}22 05 13.524  & $-$00 04 27.77  &    7.69  &  77.50  &  152  &   8.13  &  82.85  &  152  &  11.53  &  15.05  & 1546  &  1.33  &  1.96  &  CL  &  CL  \\	
   \phm{*}22 06 00.185  & +00 22 15.44  &    1.23  &   0.73  &  135  &   2.16  &   1.69  &  137  &   1.93  &   2.17  &   53  &  0.78  &  1.76  & PS  & PS  \\
   \phm{*}22 06 13.125  & +00 42 24.45  &    3.61  &   4.28  &  148  &   3.36  &   3.23  &  146  &   4.13  &   5.72  &   57  &  1.33  &  1.70  & PS  & PS  \\
   \phm{*}22 08 15.086  & +00 42 53.60  &    2.89  &   2.79  &  150  &   3.85  &   3.56  &  153  &   4.00  &   5.45  &   69  &  1.41  &  1.89  & PS  & PS  \\
   \**22 08 22.892  & +00 23 53.05  &    2.42  &   2.00  &  145  &   4.76  &   4.69  &  147  &   3.32  &   4.20  &   58  &  1.20  &  1.97  & PS  & PS  \\  
   \phm{*}22 09 22.864  & +00 28 45.23  &    2.11  &   1.70  &  138  &   2.95  &   2.39  &  139  &   3.48  &   3.58  &   58  &  0.53  &  1.70  & PS  & PS  \\
 \phm{*}22 16 24.950  & $-$00 10 10.22  &    5.11  &   5.11  &  131  &   4.44  &   4.49  &  134  &   6.21  &   6.22  &   52  &  0.09  &  1.40  & PS  & PS  \\
   \**22 20 36.320  & +00 33 34.17  &   12.76  &  12.68  &  146  &  15.54  &  15.30  &  148  &  21.13  &  25.29  &  116  &  1.00  &  1.98  & PS  & PS  \\  
   \phm{*}22 23 40.593  & +00 01 37.16  &    1.67  &   1.08  &  136  &   2.45  &   2.13  &  136  &   2.62  &   2.67  &   53  &  0.48  &  1.60  & PS  & PS  \\
 \phm{*}22 25 30.691  & $-$00 10 31.01  &    2.70  &   2.79  &  139  &   4.05  &   4.12  &  139  &   3.58  &   3.64  &   67  &  0.30  &  1.50  & PS  & PS  \\
   \phm{*}22 25 45.603  & +00 40 37.78  &    1.89  &   1.72  &  134  &   2.55  &   1.78  &  135  &   2.92  &   2.90  &   72  &  0.30  &  1.53  & PS  & PS  \\
   \**22 27 04.246  & +00 45 17.54  &    5.57  &   5.80  &  143  &   8.32  &   8.90  &  138  &  12.58  &  13.41  &  343  &  0.54  &  2.41  & PS  & PS  \\   
 \phm{*}22 27 23.275  & $-$00 25 35.82  &    4.37  &   4.10  &  138  &   3.56  &   2.60  &  141  &   4.04  &   4.68  &   58  &  0.71  &  1.31  & PS  & PS  \\
   \**22 27 26.543  & +00 10 59.25  &    4.52  &   4.72  &  154  &   6.21  &   6.24  &  154  &   4.80  &   6.79  &  127  &  1.26  &  1.50  & PS  & PS  \\  
   \phm{*}22 29 09.760  & +00 04 44.03  &    4.66  &   3.81  &  153  &   4.82  &   4.44  &  152  &   4.60  &   6.29  &   60  &  1.53  &  1.35  & PS  & PS  \\
\phm{*}22 30 19.198  & +00 50 46.84  &    1.03  &   0.94  &  137  &   1.79  &   1.95  &  136  &   1.96  &   2.06  &  192  &  0.54  &  2.00  &   PS  &  PS  \\
   \**22 30 47.467  & +00 27 56.61  &   13.82  &  14.50  &  151  &   8.73  &   9.97  &  149  &   9.41  &  13.00  &   66  &  1.32  &  1.58  & PS  & PS  \\  
   \phm{*}22 33 17.849  & +00 34 38.87  &    2.15  &   1.60  &  144  &   1.46  &   1.36  &  145  &   1.95  &   2.50  &   58  &  0.96  &  1.72  & PS  & PS  \\
   \phm{*}22 33 24.886  & +00 09 33.30  &    3.58  &   3.71  &  150  &   3.54  &   3.19  &  150  &   3.61  &   4.88  &   62  &  1.26  &  1.38  & PS  & PS  \\
 \phm{*}22 37 24.664  & $-$00 15 25.53  &    2.11  &   2.25  &  138  &   2.82  &   2.35  &  138  &   3.54  &   3.45  &   53  &  0.38  &  1.64  & PS  & PS  \\
   \phm{*}22 37 30.754  & +00 31 07.75  &    3.90  &   3.94  &  140  &   3.20  &   2.96  &  139  &   5.49  &   6.42  &   75  &  0.87  &  2.01  & PS  & PS  \\
   \phm{*}22 40 39.443  & +00 21 04.30  &    5.81  &   6.28  &  142  &   5.72  &   5.60  &  144  &   6.62  &   7.49  &   54  &  0.80  &  1.31  & PS  & PS  \\
 \phm{*}22 44 32.256  & $-$00 20 35.83  &    4.39  &   3.92  &  145  &   3.55  &   2.78  &  146  &   3.87  &   5.17  &   66  &  1.24  &  1.46  & PS  & PS  \\
 \**22 44 48.100  & $-$00 06 19.65  &    5.58  &   5.43  &  141  &   8.43  &   8.19  &  140  &   7.21  &   8.94  &   67  &  0.91  &  1.60  & PS  & PS  \\  
   \phm{*}22 46 08.961  & +00 32 14.43  &    6.09  &   5.76  &  143  &   5.61  &   5.58  &  142  &   5.44  &   7.38  &   58  &  1.08  &  1.31  & PS  & PS  \\
   \phm{*}22 46 11.896  & +00 32 32.82  &    1.02  &   1.81  &  141  &   1.43  &   0.73  &  140  &   1.49  &   1.98  &   57  &  1.04  &  1.94  & PS  & PS  \\
   \phm{*}22 46 24.124  & +00 46 24.49  &    3.65  &   3.38  &  140  &   2.92  &   2.75  &  150  &   3.69  &   4.16  &   57  &  0.81  &  1.42  & PS  & PS  \\
 \**22 46 27.685  & $-$00 12 14.19  &   56.00  &  56.81  &  136  &  86.63  &  86.60  &  136  & 100.88  & 102.84  &  652  &  0.25  &  1.84  & PS  & PS  \\  
   \**22 47 30.195  & +00 00 06.42  &  183.71  & 190.56  &  141  & 470.00  & 477.80  &  140  & 397.62  & 460.31  & 2475  &  0.79  &  2.56  & PS  & PS  \\  
   \phm{*}22 48 04.840  & +00 32 52.42  &    3.95  &   4.25  &  139  &   5.04  &   4.93  &  140  &   4.49  &   5.14  &   53  &  0.73  &  1.30  & PS  & PS  \\
   \**22 49 22.295  & +00 18 04.51  &    8.51  &   8.31  &  134  &  10.62  &  10.22  &  138  &  11.51  &  11.57  &   59  &  0.35  &  1.36  & PS  & PS  \\  
 \phm{*}22 55 37.623  & $-$00 01 44.13  &    6.60  &   6.04  &  155  &   7.34  &   7.13  &  165  &   6.74  &   8.92  &  126  &  1.36  &  1.35  & PS  & PS  \\
 \phm{*}22 59 33.637  & $-$00 28 22.60  &    1.97  &   1.24  &  141  &   1.95  &   2.22  &  148  &   2.25  &   2.88  &   55  &  1.02  &  1.48  & PS  & PS  \\
   \phm{*}23 00 02.369  & +00 00 46.48  &    1.41  &   2.25  &  134  &   1.74  &   1.61  &  136  &   1.90  &   2.44  &   51  &  1.27  &  1.73  & CL  & CL  \\
 \phm{*}23 00 08.229  & $-$00 28 16.88  &    4.10  &   3.95  &  136  &   3.68  &   3.10  &  136  &   4.01  &   5.24  &   54  &  1.23  &  1.42  & PS  & PS  \\
 \phm{*}23 01 34.862  & $-$00 20 15.94  &    1.73  &   1.23  &  139  &   2.44  &   2.22  &  150  &   2.23  &   2.82  &   54  &  0.95  &  1.63  & PS  & PS  \\
 \phm{*}23 03 10.630  & $-$00 01 34.97  &    2.81  &   2.61  &  137  &   2.64  &   2.44  &  136  &   3.04  &   3.73  &   50  &  0.98  &  1.41  & PS  & PS  \\
   \**23 03 14.818  & +00 00 52.34  &    1.72  &   1.29  &  136  &   3.85  &   3.36  &  135  &   3.12  &   3.69  &   51  &  0.78  &  2.24  & PS  & PS  \\  
 \phm{*}23 04 23.074  & $-$00 04 17.34  &    2.87  &   2.78  &  142  &   2.15  &   2.66  &  144  &   3.84  &   5.10  &   80  &  1.21  &  2.37  & PS  & PS  \\
   \phm{*}23 13 09.856  & +00 08 05.58  &    2.71  &   2.26  &  137  &   3.46  &   3.28  &  140  &   2.88  &   3.90  &   50  &  1.08  &  1.44  & PS  & PS  \\
 \phm{*}23 13 17.125  & $-$00 12 19.84  &    1.04  &   1.17  &  135  &   1.40  &   0.72  &  136  &   2.29  &   2.73  &   51  &  0.80  &  2.62  & PS  & PS  \\
   \phm{*}23 13 21.460  & +00 38 12.34  &    2.87  &   2.82  &  135  &   3.73  &   3.78  &  133  &   3.51  &   3.84  &   57  &  0.58  &  1.34  & PS  & PS  \\
 \phm{*}23 13 22.737  & $-$00 02 12.97  &    7.21  &   7.02  &  148  &   7.76  &   7.56  &  150  &   7.70  &  10.34  &   55  &  1.38  &  1.43  & PS  & PS  \\
 \phm{*}23 14 58.458  & $-$00 22 31.80  &    1.19  &   1.15  &  136  &   1.92  &   1.65  &  134  &   1.91  &   2.52  &   56  &  1.07  &  2.12  & PS  & PS  \\
 \phm{*}23 15 39.664  & $-$00 01 11.07  &    3.43  &   3.56  &  145  &   3.65  &   3.58  &  144  &   3.43  &   4.46  &   62  &  1.13  &  1.30  & PS  & PS  \\
   \**23 15 41.657  & +00 29 36.59  &   13.94  &  16.32  &  150  &  22.28  &  24.16  &  150  &  17.03  &  23.85  &   59  &  1.53  &  1.71  & PS  & CJ  \\  
   \phm{*}23 15 48.190  & +00 07 21.33  &    6.05  &   6.32  &  148  &   5.99  &   6.00  &  147  &   5.54  &   7.80  &   64  &  1.52  &  1.30  & PS  & PS  \\
 \**23 15 58.666  & $-$00 12 05.44  &    6.27  &   6.09  &  141  &   3.65  &   4.08  &  141  &   4.21  &   4.98  &   54  &  0.99  &  1.72  & PS  & PS  \\  
   \phm{*}23 16 07.791  & +00 31 07.72  &    6.00  &   5.79  &  146  &   4.96  &   4.29  &  145  &   4.67  &   6.54  &   58  &  1.14  &  1.32  & PS  & PS  \\
   \phm{*}23 18 10.992  & +00 17 49.85  &    2.26  &   2.13  &  143  &   3.35  &   2.97  &  142  &   2.53  &   3.54  &   55  &  1.27  &  1.57  & PS  & PS  \\
   \phm{*}23 18 24.423  & +00 30 51.88  &    3.82  &   3.76  &  142  &   2.81  &   2.95  &  141  &   3.26  &   3.84  &   54  &  0.83  &  1.37  & PS  & PS  \\
   \phm{*}23 20 11.604  & +00 12 19.67  &    3.32  &   2.94  &  147  &   2.96  &   2.45  &  146  &   3.45  &   4.24  &   70  &  1.21  &  1.43  & PS  & PS  \\   
 \end{longtable}
\clearpage

\clearpage

\tabletypesize{\scriptsize}
\begin{longtable}{llrrrrrrrrrrrcc}
\caption{Epoch I/II Variable Sources} \label{tab:FZDvariables} \\

\hline\hline \\[0.5ex]
       & &  \multicolumn{3}{c}{\textbf{Epoch I}} & \multicolumn{3}{c}{\textbf{Epoch II}} & \multicolumn{3}{c}{\textbf{Epoch III}} &  &  &  & \\
      \multicolumn{1}{c}{\textbf{RA}} & \multicolumn{1}{c}{\textbf{Dec}} & \multicolumn{1}{c}{\boldmath{$S_{\rm pk}$}} & \multicolumn{1}{c}{\boldmath{$S_{\rm int}$}} & \multicolumn{1}{c}{\textbf{rms}} & \multicolumn{1}{c}{\boldmath{$S_{\rm pk}$}} & \multicolumn{1}{c}{\boldmath{$S_{\rm int}$}} & \multicolumn{1}{c}{\textbf{rms}} & \multicolumn{1}{c}{\boldmath{$S_{\rm pk}$}} & \multicolumn{1}{c}{\boldmath{$S_{\rm int}$}} & \multicolumn{1}{c}{\textbf{rms}} & \multicolumn{1}{c}{\textbf{Maj}} & \multicolumn{1}{c}{\boldmath{$f_{\rm var}$}} & \multicolumn{2}{c}{\textbf{Morphology}} \\
       & & \multicolumn{1}{c}{\textbf{(mJy)}} & \multicolumn{1}{c}{\textbf{(mJy)}} & \multicolumn{1}{c}{\textbf{($\mu$Jy)}} & \multicolumn{1}{c}{\textbf{(mJy)}} & \multicolumn{1}{c}{\textbf{(mJy)}} & \multicolumn{1}{c}{\textbf{($\mu$Jy)}} & \multicolumn{1}{c}{\textbf{(mJy)}} & \multicolumn{1}{c}{\textbf{(mJy)}} & \multicolumn{1}{c}{\textbf{($\mu$Jy)}} & \multicolumn{1}{c}{\boldmath{($^{\prime\prime}$)}} & & \multicolumn{1}{c}{\textbf{(FIRST)}} & \multicolumn{1}{c}{\textbf{(S82)}}\\     [0.5ex] \hline
       \\[0.5ex]
\endfirsthead

\\[0.2ex] \hline \hline
\endlastfoot
  
     \phm{*}00 43 32.712  & +00 24 59.84  &  108.53  &    111.57  &     153  &    122.43  &    125.68  &     151  &     76.14  &    104.13  &     698  &      1.44  &    1.19  & PS & PS \\
     \phm{*}00 48 19.124  & +00 14 57.13  &   89.40  &     91.66  &     150  &    100.00  &    101.17  &     153  &     74.97  &    102.69  &     549  &      1.27  &    1.12  & PS & PS \\
     \phm{*}00 52 05.568  & +00 35 38.11  &   81.46  &     87.95  &     140  &    35.07   &    40.46  &     142  &     25.54  &     32.51  &     480  &      1.15  &    2.71  & CJ & CL \\
     \phm{*}00 52 12.473  & +00 09 45.22  &    9.77  &     10.45  &     138  &     12.15  &     12.20  &     141  &     11.43  &     11.96  &      50  &      0.46  &    1.15  & PS & PS \\
     \phm{*}00 52 25.662  & +00 26 27.99  &    9.98  &      9.77  &     154  &     12.36  &     13.45  &     155  &      8.47  &     10.93  &      60  &      1.43  &    1.35  & PS & PS \\
   \phm{*}00 57 17.004  & $-$00 24 33.26  &  114.74  &    119.63  &     142  &     90.60  &     95.62  &     141  &     82.71  &     89.16  &     890  &      0.53  &    1.34  & CJ & CJ \\ 
     \phm{*}01 05 25.524  & +00 11 21.57  &   41.73  &     44.32  &     152  &     34.46  &     35.29  &     154  &     41.52  &     63.90  &     218  &      1.47  &    1.84  & PS & PS \\
     \phm{*}01 07 45.229  & +00 39 52.89  &    8.88  &      8.84  &     137  &     10.71  &     10.49  &     137  &     10.02  &     10.93  &      53  &      0.58  &    1.24  & PS & PS \\
     \phm{*}01 08 38.569  & +00 28 14.37  &    3.91  &      4.40  &     139  &      5.48  &      5.64  &     138  &      2.47  &      3.79  &      54  &      1.45  &    1.45  & PS & PS \\
     \**01 11 06.789  & +00 08 46.51  &    3.43  &      3.67  &     147  &     5.40  &      4.99  &     149  &      3.80  &      5.09  &      67  &      1.37  &    1.39  & PS & PS \\
     \phm{*}01 15 15.786  & +00 12 48.51  &   43.13  &     45.01  &     152  &     47.47  &     48.62  &     153  &     33.48  &     45.99  &     268  &      1.36  &    1.07  & PS & PS \\
   \phm{*}01 22 13.919  & $-$00 18 01.03  &  331.60  &    348.28  &     236  &    391.18  &    404.79  &     174  &    331.27  &    405.04  &    2930  &      1.12  &    1.16  & CL & CL \\ 
   \phm{*}01 25 28.846  & $-$00 05 55.89  & 1481.35  &   1524.09  &     778  &   1349.32  &   1360.75  &     250  &   1128.81  &   1241.65  &   11014  &      0.67  &    1.23  & PS & PS \\
     \phm{*}01 27 53.705  & +00 25 16.66  &   90.08  &     92.65  &     155  &    132.93  &    135.40  &     156  &     93.00  &    140.20  &     606  &      1.35  &    1.51  & PS & PS \\
     \phm{*}01 34 57.423  & +00 39 43.06  &    2.87  &      2.73  &     139  &      6.25  &      6.38  &     146  &      1.87  &      2.37  &      74  &      1.14  &    2.62  & PS & PS \\
     \phm{*}01 38 14.968  & +00 14 44.47  &   51.76  &     62.21  &     150  &     43.57  &     92.45  &     149  &     18.91  &     62.13  &     539  &      3.81  &    1.47  & CJ & CJ \\ 
   \phm{*}01 53 29.761  & $-$00 22 14.39  &   14.98  &     15.31  &     143  &     17.75  &     18.13  &     144  &     16.57  &     18.67  &      66  &      0.79  &    1.22  & PS & PS \\
     \phm{*}01 55 28.483  & +00 12 04.55  &   19.82  &     19.92  &     134  &     17.52  &     18.16  &     133  &     18.17  &     18.40  &      55  &      0.25  &    1.12  & PS & PS \\
   \phm{*}01 59 50.086  & $-$00 24 07.14  &   12.63  &     12.63  &     134  &     10.89  &     11.01  &     133  &     11.52  &     12.25  &      55  &      0.60  &    1.17  & PS & PS \\  
     \phm{*}02 01 41.046  & +00 38 25.61  &    6.35  &      6.59  &     133  &      4.64  &      3.88  &     136  &      5.05  &      5.38  &      52  &      0.49  &    1.76  & PS & PS \\
   \phm{*}02 02 14.291  & $-$00 17 48.29  &   60.00  &     63.70  &     151  &     76.08  &     80.05  &     153  &     37.32  &     53.34  &     433  &      1.51  &    1.48  & PS & PS \\
     \**02 02 34.322  & +00 03 01.83  &   39.41  &     44.01  &     141  &     30.99  &     35.92  &     142  &     41.74  &     44.64  &     436  &      0.59  &    1.26  & CL & CL \\
   \phm{*}02 09 28.853  & $-$00 12 24.67  &    2.51  &      2.57  &     154  &      4.24  &      3.93  &     155  &      2.99  &      4.35  &      66  &      1.36  &    1.69  & PS & PS \\
   \phm{*}02 12 02.134  & $-$00 27 49.97  &   52.84  &     55.28  &     134  &     45.75  &     47.14  &     133  &     46.07  &     47.65  &     483  &      0.36  &    1.19  & CL & CL \\
   \phm{*}02 13 01.143  & $-$00 18 14.97  &   48.44  &     52.80  &     145  &     42.12  &     46.56  &     150  &     35.43  &     48.18  &     349  &      1.41  &    1.15  & PS & CJ \\
     \phm{*}02 15 53.647  & +00 18 26.92  &   35.82  &     37.76  &     148  &     31.35  &     33.33  &     148  &     28.46  &     38.48  &     241  &      1.37  &    1.17  & PS & CJ \\
   \phm{*}02 17 56.005  & $-$00 09 36.26  &    5.85  &      7.98  &     149  &      4.10  &      6.12  &     148  &      2.38  &      6.70  &      90  &      3.25  &    1.34  & CJ & CJ \\  
   \phm{*}02 18 40.536  & $-$00 15 16.22  &   13.17  &     14.12  &     153  &     10.27  &     10.92  &     153  &      7.66  &     10.63  &      73  &      1.36  &    1.33  & CX & PS \\
   \phm{*}22 07 55.252  & $-$00 02 15.07  &   61.93  &     68.63  &     153  &     79.40  &     85.23  &     152  &     47.30  &     67.57  &     433  &      1.30  &    1.25  & CJ & CJ \\   
     \**22 08 22.892  & +00 23 53.05  &    2.42  &      2.00  &     145  &      4.76  &      4.84  &     147  &      3.32  &      4.20  &      58  &      1.20  &    2.35  & PS & PS \\
   \phm{*}22 10 01.826  & $-$00 13 09.76  &  115.31  &    118.15  &     139  &    127.38  &    132.29  &     141  &    108.43  &    123.59  &     938  &      0.75  &    1.11  & PS & PS \\
   \phm{*}22 10 31.474  & $-$00 13 55.93  &   12.00  &     12.17  &     134  &     14.31  &     14.61  &     136  &     14.40  &     14.79  &      70  &      0.32  &    1.22  & PS & PS \\
     \phm{*}22 19 09.385  & +00 31 12.64  &   11.70  &     16.43  &     146  &      9.33  &     13.66  &     147  &      6.56  &      8.05  &      73  &      1.06  &    2.04  & CJ & CJ \\  
     \**22 20 36.320  & +00 33 34.17  &   12.76  &     12.68  &     146  &     15.54  &     15.57  &     148  &     21.13  &     25.29  &     116  &      1.00  &    1.99  & PS & PS \\
   \phm{*}22 21 35.004  & $-$00 11 00.18  &    8.64  &      8.74  &     151  &     14.36  &     14.30  &     153  &      7.07  &     10.18  &      90  &      1.43  &    1.61  & PS & PS \\
     \phm{*}22 22 35.856  & +00 15 36.36  &   51.13  &     61.44  &     133  &     46.67  &     55.14  &     132  &     34.75  &     42.85  &     498  &      1.04  &    1.43  & CJ & CJ \\
     \**22 27 04.246  & +00 45 17.54  &    5.57  &      5.80  &     143  &      8.32  &      9.09  &     138  &     12.58  &     13.41  &     343  &      0.54  &    2.31  & PS & PS \\
     \**22 27 26.543  & +00 10 59.25  &    4.52  &      4.72  &     154  &      6.21  &      6.40  &     154  &      4.80  &      6.79  &     127  &      1.26  &    1.44  & PS & PS \\
     \phm{*}22 27 29.071  & +00 05 22.07  &   91.64  &     97.51  &     148  &     78.83  &     84.02  &     148  &     61.52  &     81.50  &     651  &      1.39  &    1.20  & CL & CX? \\ 
     \phm{*}22 27 44.589  & +00 34 50.90  &   30.53  &     31.64  &     154  &     22.48  &     23.57  &     153  &     18.19  &     23.77  &     188  &      1.40  &    1.36  & PS & PS \\
     \phm{*}22 27 58.134  & +00 37 05.46  &   99.12  &    103.79  &     146  &     69.58  &     72.61  &     145  &     64.45  &     68.32  &     632  &      0.73  &    1.52  & PS & PS \\
     \**22 30 47.467  & +00 27 56.61  &   13.82  &     14.50  &     151  &      8.73  &     10.18  &     149  &      9.41  &     13.00  &      66  &      1.32  &    1.45  & PS & PS \\
     \phm{*}22 42 24.141  & +00 55 13.41  &   13.44  &     16.08  &     139  &     18.01  &     21.18  &     145  &     13.44  &     16.08  &     212  &      0.88  &    1.30  & CJ & CJ \\  
   \phm{*}22 43 31.930  & $-$00 12 33.08  &   21.40  &     21.67  &     133  &     14.10  &     14.80  &     134  &     14.64  &     15.33  &      81  &      0.44  &    1.49  & PS & PS \\  
   \**22 44 48.100  & $-$00 06 19.65  &    5.58  &      5.43  &     141  &      8.43  &      8.38  &     140  &      7.21  &      8.94  &      67  &      0.91  &    1.65  & PS & PS \\
   \**22 46 27.685  & $-$00 12 14.19  &   56.00  &     56.81  &     136  &     86.63  &     87.70  &     136  &    100.88  &    102.84  &     652  &      0.25  &    1.81  & PS & PS \\
     \**22 47 30.195  & +00 00 06.42  &  183.71  &    190.56  &     141  &    470.00  &    483.43  &     140  &    397.62  &    460.31  &    2475  &      0.79  &    2.51  & PS & PS \\
     \**22 49 22.295  & +00 18 04.51  &    8.51  &      8.31  &     134  &     10.62  &     10.43  &     138  &     11.51  &     11.57  &      59  &      0.35  &    1.39  & PS & PS \\
   \phm{*}22 58 52.939  & $-$00 18 57.31  &    --      &      3.45  &     143  &     --       &      1.28  &     149  &    --        &      0.25  &    --      &      3.80  &    --      & PS & $-$ \\
     \phm{*}23 01 57.804  & +00 03 51.81  &    7.04  &      7.24  &     149  &      3.88  &      3.94  &     149  &      5.02  &      7.56  &      67  &      1.38  &    1.99  & PS & PS \\
     \**23 03 14.818  & +00 00 52.34  &    1.72  &      1.29  &     136  &      3.85  &      3.49  &     135  &      3.12  &      3.69  &      51  &      0.78  &    2.86  & PS & PS \\
     \phm{*}23 06 55.166  & +00 36 38.11  &   15.19  &     15.39  &     142  &     13.23  &     13.00  &     141  &     12.81  &     16.17  &     113  &      1.13  &    1.27  & PS & PS \\
     \**23 15 41.657  & +00 29 36.59  &   13.94  &     16.32  &     150  &     22.28  &     24.53  &     150  &     17.03  &     23.85  &      59  &      1.53  &    1.48  & PS & CJ \\ 
   \**23 15 58.666  & $-$00 12 05.44  &    6.27  &      6.09  &     141  &      3.65  &      4.22  &     141  &      4.21  &      4.98  &      54  &      0.99  &    1.49  & PS & PS \\
   \phm{*}23 18 45.827  & $-$00 07 54.91  &    5.34  &      5.13  &     149  &      3.16  &      3.41  &     151  &      3.14  &      4.68  &      56  &      1.59  &    1.56  & PS & PS \\
     \phm{*}23 18 56.659  & +00 14 37.97  &   18.34  &     20.18  &     142  &     20.48  &     22.40  &     143  &     17.84  &     25.95  &      56  &      1.33  &    1.29  & PS & PS \\ 
     \phm{*}23 19 10.344  & +00 18 59.10  &   33.36  &     33.09  &     141  &     29.33  &     29.90  &     141  &     30.32  &     35.19  &     234  &      0.78  &    1.19  & PS & PS \\
     \phm{*}23 20 38.001  & +00 31 39.47  &   82.95  &     86.04  &     138  &     73.62  &     74.94  &     140  &     33.09  &     52.30  &     724  &      1.69  &    1.65  & PS & PS \\
\end{longtable}
\clearpage

\begin{table}[t]
  \centering
  \caption{Fractional Variability}
  \begin{tabular} {lrrr}
       \hline\hline \\[-2ex]
       \textbf{Fractional Variability} & \textbf{This Work} & \textbf{de Vries} & \textbf{Galactic} \\      
       \textbf{f} 				  & \textbf{N}  		 & \textbf{N}  	      & \textbf{N} \\  [0.5ex] \hline
         \\[-1.8ex]
       $<$ 1.25   & -   & 51   & 1  \\
       1.25 - 1.50   & 39  & 39   & 5 \\
       1.50 - 1.75   & 23  & 19   & 5 \\
       1.75 - 2.00   & 15  & 4 	  & 4 \\
       2.00 - 2.25   & 6   & 4    & 2 \\
       2.25 - 2.50   & 3   & 2    & 3 \\
       2.50 - 2.75   & 2   & 1    & 2 \\
       2.75 - 3.00   & 0   & 1    & 0 \\
       $>$ 3.0 	     & 1   & 2    & 17 \\ [0.5ex] \hline
       \\[-1.8ex]
       Total 	     & 89 & 123  & 39 \\
     \\[-1.8ex]  \hline\hline
  \end{tabular}
  \label{tab:fvar}
\end{table}

\begin{table*}
  \centering
  \caption{Optical Identifications}
  \begin{tabular} {lccc}
       \hline\hline \\[-2ex]
        	  & \textbf{SDSS Match Rate} & \textbf{Galaxy Fraction} & \textbf{Stellar Fraction} \\ [0.5ex]\hline
	\\[-1.8ex]			    
        Variable Sources  & 60\% $\pm$ 5\%  & 72\% $\pm$ 6\%  & 28\% $\pm$ 6\% \\   
        Non--variable Control & 44\% $\pm$ 5\%  & 92\% $\pm$ 4\%  & 8\% $\pm$ 4\% \\
       \\[-1.8ex] \hline\hline
  \end{tabular}
  \label{tab:sdss}
\end{table*}

\begin{table*}
  \centering
  \caption{Quasar Fraction}   
    \begin{tabular} {lcc}
       \hline\hline \\[-2ex]
        		& \textbf{Spectroscopic QSO} 	& \textbf{Photometric QSO} \\[0.5ex]\hline
	\\[-1.8ex]						
       Number sources in area		& 2083			& 6986 \\
       Variable Source QSO Fraction 	& 10\% $\pm$ 3\%	& 13\% $\pm$ 4\% \\
       Control Sample QSO Fraction     	& 3\% $\pm$ 2\%		& 2\% $\pm$ 2\% \\          
    \\[-1.8ex]   \hline\hline
  \end{tabular}
  \label{tab:qsos}
\end{table*}

\end{document}